%% file: main.tex
\documentclass[sigconf, nonacm]{acmart}

\input{commands}

\AtBeginDocument{%
  }

\begin{document}

\title{\sysname{}: Editable Planning for Controllable Image Generation}

\author{Susie S.Y. Li}
\orcid{0009-0000-6375-182X}
\authornote{Work done during and internship at Apple.}
\correspondingauthor
\affiliation{%
  \institution{Tufts University}
  \city{Medford}
  \state{MA}
  \country{USA}
}
\email{suyang.li@tufts.edu}

\author{Angie Boggust}
\orcid{0000-0002-9671-5574}
\affiliation{%
  \institution{Apple}
  \city{Cambridge}
  \state{MA}
  \country{USA}
  }
\email{aboggust@apple.com}

\author{Fred Hohman}
\orcid{0000-0002-4164-844X}
\affiliation{%
  \institution{Apple}
  \city{Seattle}
  \state{WA}
  \country{USA}
  }
\email{fredhohman@apple.com}

\author{Donghao Ren}
\orcid{0000-0001-8666-7241}
\affiliation{%
  \institution{Apple}
  \city{Seattle}
  \state{WA}
  \country{USA}
  }
\email{donghao@apple.com}

\author{Dominik Moritz}
\orcid{0000-0002-3110-1053}
\affiliation{%
  \institution{Apple}
  \city{Pittsburgh}
  \state{PA}
  \country{USA}
  }
\email{domoritz@apple.com}

\author{Yannick Assogba}
\orcid{0000-0002-6646-0961}
\correspondingauthor
\affiliation{%
  \institution{Apple}
  \city{Cambridge}
  \state{MA}
  \country{USA}
  }
\email{yassogba@apple.com}

\input{sections/00_abstract}

\begin{teaserfigure}
 \centering
 \includegraphics[width=\textwidth]{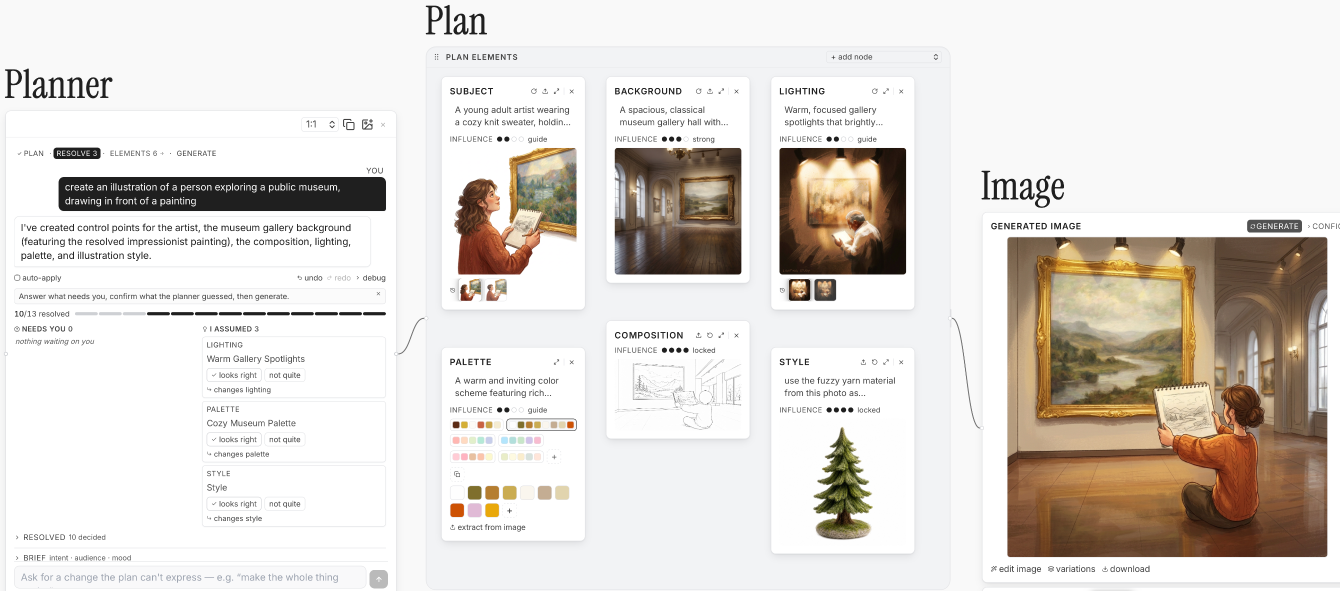}
  \caption{Instead of a black-box prompt-to-image interface, \sysname provides a planning agent (left) that collaborates with creators to clarify user intent and surface model assumptions. The agent then decomposes the image generation goal into semantically relevant design decisions represented as plan elements (middle). The user can freely modify the plan elements, add their own visual and textual references after which \sysname synthesizes a prompt to capture this rich user specification to generate the final image (right).}
  \Description{Three-stage reComposer workflow from planning to final generation. The left panel shows a planning agent that clarifies intent and surfaces assumptions; the center shows editable plan elements for decisions such as subject, background, palette, composition, and style; the right shows the final image synthesized from the plan.}
  \label{fig:teaser}
\end{teaserfigure}

\maketitle

\input{sections/01_introduction}
\input{sections/02_related_work}
\input{sections/03_formative}
\input{sections/04_system_description}
\input{sections/05_evaluation_and_results}

\input{sections/06_discussion_and_conclusion}

\bibliographystyle{ACM-Reference-Format}
\bibliography{main}

\onecolumn
\appendix
\input{sections/21_appendix_b}
\input{sections/22_appendix_c}

\end{document}

%% file: commands.tex
\usepackage{xspace}
\usepackage{xcolor}
\usepackage{tabularx}
\usepackage{float}
\usepackage{subcaption}
\usepackage[capitalize]{cleveref}
\usepackage{booktabs}
\usepackage{natbib}
\usepackage{tikz}
\usepackage{listings}

\lstdefinestyle{prompt}{%
  basicstyle=\ttfamily\footnotesize,
  breaklines=true, breakatwhitespace=false,
  columns=fullflexible, keepspaces=true, showstringspaces=false,
  frame=single, framesep=6pt, rulecolor=\color{gray!60},
  backgroundcolor=\color{gray!7}, xleftmargin=8pt, xrightmargin=8pt,
  aboveskip=8pt, belowskip=8pt,
}

\usetikzlibrary{positioning, arrows.meta, calc}

\newcommand{\cjkimg}[1]{\raisebox{-0.15em}{\includegraphics[height=1em]{figures/#1}}}

\newcommand{\sysname}{\textsc{reComposer}\xspace}

\newcommand{\numparticipants}{8\xspace}

\newcommand{\pquote}[1]{``\textit{#1}''}

\newcommand{\genaibench}{GenAI-Bench\xspace}
\newcommand{\qwenimagebench}{Qwen-Image-Bench\xspace}

\newcommand{\raw}{\textsc{raw}\xspace}
\newcommand{\plan}{\textsc{plan}\xspace}

%% file: sections/00_abstract.tex
\begin{abstract}
    Prompt-based image generation makes it easy to produce visual assets, but offers little support for engaging with the creative process. Prompting collapses interdependent creative decisions into a monolithic instruction and returns a finished image, leaving control limited to revising the prompt or correcting the model’s output. We present \sysname, a process-oriented image-generation interface built around editable planning, which decomposes users’ evolving intent into meaningful creative decisions – such as subject, composition, or palette – and collaborates with users to explore underspecified goals and alternatives. \sysname supports conversational and direct manipulation of the plan, moving between concept-level intent and visual-level control. We evaluated \sysname with 8 design and non-design professionals. Compared to prompting, participants felt greater control and ownership while developing visual assets, exploring alternatives, and iterating on details. Our findings suggest how shifting generative AI interaction from specifying outputs to shaping process can support creative thinking alongside asset production.
\end{abstract}

%% file: sections/01_introduction.tex
\section{Introduction}

Text-to-image models increasingly support visual creation across a broad range of users, including design and non-design professionals who create visual assets as part of their work.
Yet the dominant interaction remains an opaque prompt-and-output loop: users describe an image, inspect a rendered result, and revise the prompt or describe an edit when the output diverges from their intent~\cite{he2025automatedblackboxpromptengineering, 10.1145/3491102.3501825}.  Other interfaces have explored node-based workflows and direct manipulation to provide more structured control, but often requires users to construct technical workflows or focus on particular visual dimensions~\cite{ComfyUI, shi2025brickifyenablingexpressivedesign, mou2023dragondiffusionenablingdragstylemanipulation}.
The model also often implicitly makes consequential creative decisions on the user's behalf, often before the those decisions have been consciously specified by the user~\cite{desimone2026creooneshotimagegeneration, Wadinambiarachchi_2024}.

This prompt based  workflow treats creative intent as something that can be fully specified and communicated in text upfront. Visual design and creation, however, develops through progressive exploration of interdependent decisions such as subject, color palette, composition, and style~\cite{10.1145/3706599.3720182}. Prompt-based generation compresses this emergent process into a single upfront specification. For creative tools, this means supporting asset production without necessarily supporting the process of developing the intent behind it.

Additionally, these decisions can be difficult to both express and inspect in prompt-based interfaces. Nuanced visual intentions must be translated into language, often requiring users to find the right textual description for a concept, while the model's interpretation remains largely hidden. Users see the resulting image, but have limited visibility into how individual aspects of their request were interpreted. Without persistent representation of these intermediate decisions, users have limited support for inspecting, revisiting, or expressing them as their intent evolves, or for changing one aspect while preserving others.

These challenges suggest an opportunity to shift from specifying a finished output toward explicitly developing the intermediate decisions that shape the output. In this work, we refer to this approach as \textit{editable planning}: the model proposes an initial decomposition of the creative task into relevant design decisions\,---\,such as subject, composition, style, and color palette\,---\,which users can inspect and revise as their intent develops. We instantiate this approach in \sysname, an interactive system for process-oriented visual co-creation with generative AI.
Given an initial prompt, \sysname's planner surfaces underspecified aspects of the request and proposes a task-specific set of plan elements representing creative decisions\,---\,such as subject, composition, style, and color palette\,---\,that guide generation. Rather than exposing low-level model operations or requiring users to carefully engineer a prompt, \sysname organizes these elements around concepts familiar to visual creators. Users can develop the plan conversationally or directly edit individual elements, progressively refining their intent while deciding which aspects to specify themselves and which to leave open to the generative model.

The resulting plan persists as a canvas of editable elements, each paired with interactions appropriate to the decisions it represents. For example, users can manipulate composition spatially, edit color palettes directly from color swatches or extracted from reference images, provide visual references for subjects, backgrounds and styles, and control how strongly each element influences the final output.
By making intermediate planning artifacts visible and editable, \sysname supports movement between developing high-level creative intent and making precise visual edits while making the process visible. Through \sysname, we investigate \textit{editable planning} as a process-oriented interaction paradigm for generative visual co-creation.

Our contributions are as follows:
\begin{itemize}
    \item \textit{Editable planning}, a process-oriented interaction paradigm for generative image workflows in which an AI system collaborates with a user to dynamically construct a task-specific representation of intermediate design decisions that serves as a shared, editable representation for developing creative intent.
    \item \sysname, a canvas-based interface that instantiates the editable planning for image generation through semantic plan elements and direct manipulation controls grounded in real-world workflows, providing granular control over both concept-level and visual-level aspects of the final image.
    \item An evaluation of the system through a user study with \numparticipants professionals, demonstrating its use case in a variety of real-world design tasks and scenarios. We find that \sysname's decomposition matched participants' existing visual design process and gave them more agency in the image generation process.
\end{itemize}

%% file: sections/02_related_work.tex
\section{Related Work}
In this section, we review prior work on creative workflows and practices, interaction paradigms for generative image creation, and planning in human-AI collaboration. Together, these areas characterize how creators develop and express visual intent, and motivate editable planning as an interaction paradigm for generative creation.

\subsection{Mismatch Between Creative Process and Current GenAI Workflow}

\paragraph{Creative workflows are iterative and nonlinear}
Creative processes almost never unfold in a single step of specification~\cite{Sawyer2021Iterative}. Instead, prior work on design and creative practices describes them as iterative processes in which creators move through different stages of activity, using different representations and forms of reasoning at each stage~\cite{designcouncil2005, Dorst2001Creativity}.
Similarly, Stanford University's Design School proposes a model that frames design as a nonlinear and iterative process, moving through five different stages~\cite{stanford}.

\paragraph{Creative workflows are practice-specific}
Although creative processes often share broad patterns of exploration and refinement, the intermediate design decisions through which this occurs vary across practices.
For example, in drawing and painting, artists commonly reason through composition, value, and color as distinct aspects of the image~\cite{ma2025artkit}.
Whereas infographic designers often progressively build a design by establishing grid structure and visual hierarchy, defining typography and color, and exploring alternative design decisions in parallel~\cite{10.1145/3706599.3720182, Kuba02042024}. Illustration practice similarly introduces domain-specific actions, such as narrative interpretation, character and composition development, sketch refinement, and layout adjustment~\cite{arifin2025illustration}.

\subsection{Interaction Paradigms for GenAI Creativity Tools}
\paragraph{Prompt-based interfaces}
Generative AI systems such as Gemini~\cite{Gemini35Flash}, Midjourney~\cite{midjourney}, and ChatGPT~\cite{openai2026chatgpt} use natural language prompts as the primary user interaction method, making image generation accessible by allowing users to express creative intent in natural language. Although prompting lowers the barrier to creating images, users often experience difficulties in achieving an output that matches their intent.
Translating nuanced visual goals into text often requires meticulous prompt engineering.
Optimal prompts are substantially more complex and lengthy than users' natural descriptions, creating a gap between how users intuitively communicate intent and the prompt-engineering required for effective outputs~\cite{he2025automatedblackboxpromptengineering, hao2023optimizingpromptstexttoimagegeneration, shi2025brickifyenablingexpressivedesign}. As prompts accumulate constraints and context, they become difficult to interpret, revise, and reuse.

\paragraph{Node-based interfaces}
In contrast, \textit{pipeline-oriented node-based interfaces} like ComfyUI~\cite{ComfyUI} expose model parameters, conditioning mechanisms, and operations as modular graphs, enabling users to compose precise and reusable generation pipelines. However, their technical complexity makes them inaccessible to many users.
To effectively leverage these systems, users must understand low-level computational components, translate creative intent into model inference operations, and configure how those operations connect~\cite{nguyen2025genflowinteractivemodularimage}.

Beyond pipeline-oriented interfaces, recent work explores \textit{intent-oriented node interfaces}, using node-based representations to organize creative processes at a higher level of abstraction. Rather than model operations, these systems represent creative artifacts, transformations, and alternative creative decisions as modular nodes in the system. For example, Spellburst~\cite{angert2023spellburst} represents generative creative code as branchable and mergeable nodes, while Kyaw et al.~\cite{kyaw2025nodebasededitingmultimodalgeneration} use node-based graphs to support composition and targeted editing of multimodal narratives in creative authoring, showing how visual-centric nodes can better support exploration while reducing the workload of manual workflow construction~\cite{nguyen2025genflowinteractivemodularimage}. Empirical findings from these systems demonstrate how \textit{modular representations} can \textit{externalize creative process and structure} and provide localized control points.

\sysname builds on this principle but shifts the unit of modularity to the semantic decisions within the visual artifact: it automatically decomposes a natural-language description into an \textit{editable plan}, allowing users to progressively explore and iterate on individual decisions of a creative workflow without authoring a computational workflow or primarily iterating on a monolithic prompt.

\paragraph{Direct manipulation interfaces} Direct manipulation systems allow users to convey creative intent by prioritizing visual and spatial interactions, including sketching layouts~\cite{zhang2024sketchguidedsceneimagegeneration,sarukkai2024blockdetailscaffoldingsketchtoimage, sketchgpt},  spatial arrangements and bounding boxes.
Recent work draws from real-world creative and design practices. For example,
Brickify~\cite{shi2025brickifyenablingexpressivedesign} enables users to compose designs by collecting and assembling reference images.
Our work builds on top of these findings and integrates direct manipulation into each modality in the workflow: in \sysname, direct manipulation and agent-driven planning address \textit{complementary levels of control}.

\subsection{Planning as an Interface/Interaction for Human-AI Co-Creation}
\paragraph{Plans provide intermediate representation before generating an output.} Prior interactive reasoning systems show that exposing intermediate representations can help users inspect and steer generation, but this interaction has been studied predominantly in text and code generation~\cite{pang2025interactivereasoningvisualizingcontrolling, zhang2023planninglargelanguagemodels, jiang2025selfplanningcodegenerationlarge}.

Planning has also emerged in visual generation, where systems first construct intermediate representations such as scene description, object list, layout, regions, or sequences of generation operations before rendering a final image~\cite{ji2026instructionbasedimageeditingplanning, he2025plangen, zhang2026thinkstrokespixelsprocessdriven}. However, these plans are primarily treated as internal model representations: they are produced and used to condition the final output, but remain hidden from users. Consequently, users have limited opportunity to inspect, revise, or interact with the intermediate decisions and assumptions that produced the final image.

Comparatively little work explores the opportunity of \textit{editable planning for visual co-creation}: translating an open-ended prompt into interpretable design decisions that users can inspect, revise, and connect to generated results.  Closest to our work is ToMigo~\cite{hegemann2026tomigointerpretabledesignconcept}, which focuses on graphic design and uses a graph representation of design elements to guide generation. However, ToMigo's nodes express design intent only via text (e.g. by interpreting the intent behind uploaded visual references and writing that out as text), whereas our system provides image previews of these intermediate artifacts and supports visual iteration on individual design decisions before generating the full image.

Our system, \sysname, treats planning as a shared authoring representation between the user and the model. It automatically decomposes an initial prompt into editable decisions about design dimensions such as subject, composition, color, style, and other visual properties. Rather than requiring users to manually specify the full workflow, the system proposes an initial set of design decisions, and allows the user to iterate on it directly or in collaboration with the planning agent to explore and refine their intent. Our system also supports directly using visual references to prompt the final image generation.

%% file: sections/03_formative.tex
\section{Formative Research and Interviews}
\label{sec:formative}

To ground our design goals in real-world creative practice, we conducted semi-structured expert interviews with three image generation practitioners at a large technology company. The interviews focused on understanding users' approaches and challenges they encounter with current image generation tools and workflows. Our experts include: \textbf{E1}, who works on an image generation system used by machine learning engineers, creative design professionals, and business experts. \textbf{E2} is an engineer working on the same system, with prior professional experience in film, art and graphics. \textbf{E3} is an instructor and former graphic designer who teaches classes on using image generation in professional contexts. This range allowed us to understand current tools from both developer and practitioner perspectives.

\subsection{Findings and Challenges}
\label{sec:findings}

\textbf{C1: Users' creative intent develops through interaction with intermediate artifacts.} Creators begin with goals that vary in specificity. For example, some begin with a primarily text-based description of the message they are designing a visual communication for, while some begin with a set of visuals that they use as reference images. Creators rarely begin with a fully specified visual outcome and instead use generated artifacts\,---\,both textual and visual\,---\,to discover and progressively develop their goal (E1, E3). For example, users often generate images from their initial prompt, and use the initial results to further prompt the model for edits.

\textbf{C2: Limited support for linked conceptual and visual reasoning.} When creating a visual asset meaningful to their work, creators often communicate intent and reason through both semantic and visual specifications. E3, an instructor who regularly teaches company-wide classes and workshops to professional users, points out the wide variety of user backgrounds and their use cases: their domains span "product prototype, marketing and advertising materials, concept sketches, and slide deck illustration". Some users have a set of reference images that they want to follow the style of, while some have a visual message they want to communicate as supporting assets in a slide deck. However, in current prompt based interfaces, there is very limited room for users to understand the model's interpretation of concepts and how they are represented visually.

\textbf{C3: Barrier in prompt-crafting results in inefficient workflows.} Many users recognize the importance of prompt-engineering in achieving the visual that matches their goal, however almost all users are unsure of their prompting strategy due to its open-endedness. Users often hesitate on \textit{how many details} to include, how to \textit{structure} them inside a prompt, and how to \textit{plan prompts across multiple turns} with the model to achieve desired results. As a result, users often resort to complex and tedious workflows such as prompting a different language model for an image prompt, using the separate model to iterate on the image artifact, which introduce inefficiencies and still often result in discarding or repeatedly starting new sessions due to the inherent linear nature of current chat dialogues and the ineffectiveness in referencing visual artifacts produced earlier in the conversation.

%% file: sections/04_system_description.tex
\section{\sysname}
To investigate how \textit{editable planning} can support human-AI co-creation in visual work, we developed \sysname, a web-based system that externalizes evolving creative intent as an editable plan and supports iterative refinement across semantic and visual representations.

\subsection{System Design Goals and Rationale}
\label{sec:goals}
Based on findings from our formative interviews as well as prior literature, we develop the following design goals:

\textbf{DG1: Externalize evolving creative intent as editable representations.} Creative intent develops through multiple interdependent decisions (e.g. style, color palette, subject, and composition) that are often left implicit in prompts or hidden in the model's interpretation (C3). Therefore, the
system should make creative intent itself an external, inspectable artifact\,---\,decomposed
into parts that a designer can read without machine-learning vocabulary,
and adapted to the kind of visual artifact being made (C3).  \textit{This requires the system to dynamically decompose creative intent into persistent, creator-facing plan elements tailored to the task and the type of visual artifact being created}.

\textbf{DG2: Preserve creator agency through selective and reversible control.} Creative collaboration with generative models requires balancing user direction with the model's ability to propose and imagine alternative possibilities. Rather than requiring users to specify every decision or delegating the entire process to the model, the system should strive for a fine-grained balance between letting users decide which creative decisions to make themselves, which to leave open to model interpretation, and allow users to selectively revise aspects of their process as their intent develops and evolves. \textit{This requires intermediate decisions to remain independently editable and support different degrees of influence over generation.}

\textbf{DG3: Support progressive refinement across levels and modalities.} Creative work moves between high-level conceptual decisions and increasingly specific visual details, and creators express intent through both language and visual representations (C2). \textit{The system should allow users to move fluidly between conceptual planning and direct visual manipulation, using representations appropriate to each decision and stage of the process.} For example, users may describe style in language or through references, manipulate composition spatially, and edit palette through colors rather than textual descriptions.

Guided by our design goals, we develop \sysname, a system that comprises
\begin{itemize}
    \item A \textbf{planning agent} that supports task refinement, decomposition of user goals into design axes, and interactive editing of an image generation plan. (DG1,3)
    \item A structured representation of editable \textbf{plan elements} that give users control over design decisions (DG2). These representations are represented as UI elements which we also refer to as \textit{nodes}, which provide specific interaction affordances particular to that design axis.
    \item A \textbf{prompt synthesizer} that converts the plan elements into a rich prompt used to generate the final image.
\end{itemize}

Our system is modular with respect to the underlying models used, in this study we use Gemini 3.5 Flash)~\cite{Gemini35Flash}) for the planning agent and Gemini 3 Pro Image~\cite{Gemini3ProImage} and Gemini 3.1 Flash Image~\cite{Gemini31FlashImage}) for image generation.

\subsubsection{Planning Agent}

The planning agent supports the user in three primary areas, \textit{task refinement and underspecification}, \textit{plan element construction}, and \textit{coordinated  edits to existing plan elements}.

\begin{figure}[t]
\includegraphics[width=0.9\linewidth]{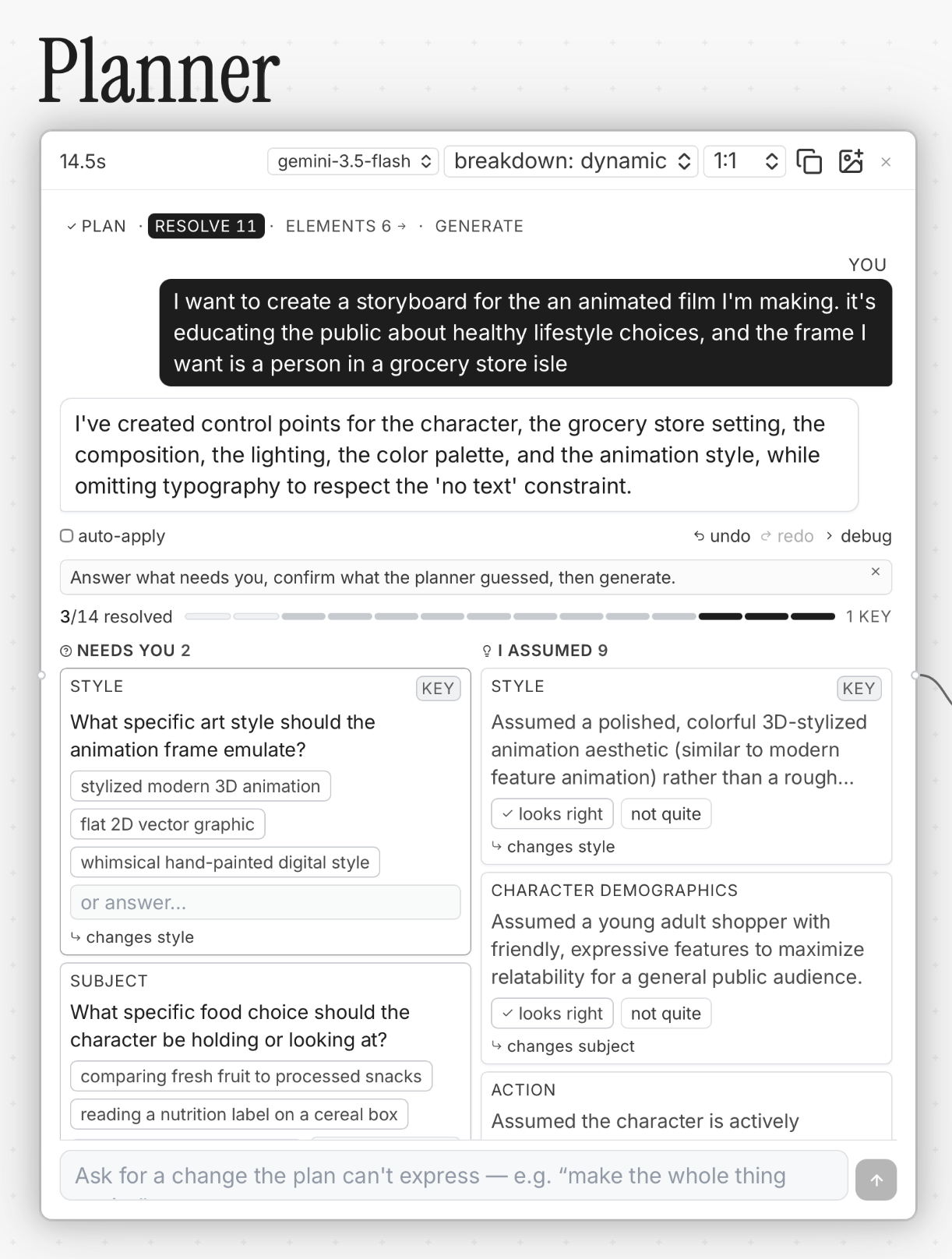}
    \caption{The planner surfaces decisions that are ambiguous in the user's prompt, as well as implicit assumptions made by the model, such as the animated character's demographic.}
    \Description{Planning agent surfacing underspecified design decisions. The interface shows a storyboard request together with unresolved questions and model assumptions about style, subject, character demographics, and action that the user can review or revise.}
    \label{fig:planner}
\end{figure}

The planning agent works with the user through a chat-style interface (\autoref{fig:planner}) to craft an image generation plan (DG3). Given an initial request, the planner identifies underspecified or ambiguous aspects of the user's intent and surfaces these through \textit{clarifying questions} and \textit{model assumptions}. The user can answer these questions, confirm or reject the assumptions, or continue describing their intent; the planner incorporates these responses into the evolving plan.

As the user interacts with the agent, the planner directly constructs and modifies the plan elements. The agent is provided the available plan element types and their functionality (\autoref{tab:node-editors}), along with some criteria for what make including a plan element useful. For example: a plan element should be meaningful for the current image task, independently editable by user, and likely to affect generation, while an ineffective control point is generic and irrelevant and may be overly redundant with another control point. From these guidelines, the planner instantiates a task-specific set of plan elements. For example, \texttt{lighting} may be useful for a photorealistic or 3D-rendered scene but omitted when creating a flat vector poster or icon. Prompts and further details are found in appendix \ref{sec:appendix-decompose-prompt}.

The planner can also perform multiple coordinated edits over the existing plan as the user's idea evolves, adding, removing, or revising multiple plan elements in response to a single conversational turn (\autoref{sec:appendix-prompt-agent}). For example, a user may request for a "softer" visual character without immediately knowing which specific visual dimensions to adjust. In these cases, the planner interprets and infers which aspects are related - such as palette, lighting, and style - and proposes a set of edits for the user to review and apply. These changes are made through the same persistent plan that users can inspect and manipulate directly, allowing broad conversational intent to resolve into explicit, revisable design decisions (DG3).

\subsubsection{Plan Elements}

\begin{table*}[ht]
\caption{\sysname's plan element types. There are 7 unique editor interfaces to support the 10 plan element types.}
\Description{Plan-element types and their editor interfaces. The ten element types are subject, composition, background, lighting, palette, style, typography, reference image, reference text, and chart. These use seven editor interfaces, including image, composition, palette, typography, text, and chart-specific editors.}
\label{tab:node-editors}
\small
\begin{tabularx}{\textwidth}{@{}l>{\raggedright\arraybackslash}X>{\raggedright\arraybackslash}X@{}}
  \textbf{Plan Element} & \textbf{Design Axis} & \textbf{Editor} \\

  \texttt{subject} & The main entity or entities: identity, appearance,
  pose. & Image editor \\
  \addlinespace

  \texttt{composition} & Arrangement, framing, and spatial layout. & Composition editor: drag/resize/reorder regions; freeform
  \emph{scene} or ordered-step \emph{graphic} layout; sketch/pen tool. \\
  \addlinespace

  \texttt{background} & The setting or backdrop behind the subject. &
  Image editor. \\
  \addlinespace

  \texttt{lighting} & Light direction, quality, and time of day. Only for
  photographic or 3D-rendered scenes. & Image editor. \\
  \addlinespace

  \texttt{palette} & The dominant and accent colors. & Palette editor: extract color palette from reference image; generate palette options from natural language prompts;  \\
  \addlinespace

  \texttt{style} & Medium and rendering aesthetic, applied globally to the
  whole image. & Image editor. \\
  \addlinespace

  \texttt{typography} & Text and font selection. & Typography editor (font, size, color). \\
  \addlinespace

  \texttt{reference image} & A user-uploaded reference image to guide generation.
  & Image editor \\
  \addlinespace

  \texttt{reference text} & User-uploaded text to include in the generated image.
  & Text editor \\
  \addlinespace

  \texttt{chart} & User uploaded chart specification. & Visualization (e.g. Vega-Lite) spec with live chart preview. \\
\end{tabularx}

\end{table*}
\label{sec:plan-elements}

\begin{figure*}[t]
  \centering
  \captionsetup[sub]{justification=raggedright,singlelinecheck=false}
  \begin{subfigure}[t]{0.35\textwidth}
    \includegraphics[width=\linewidth]{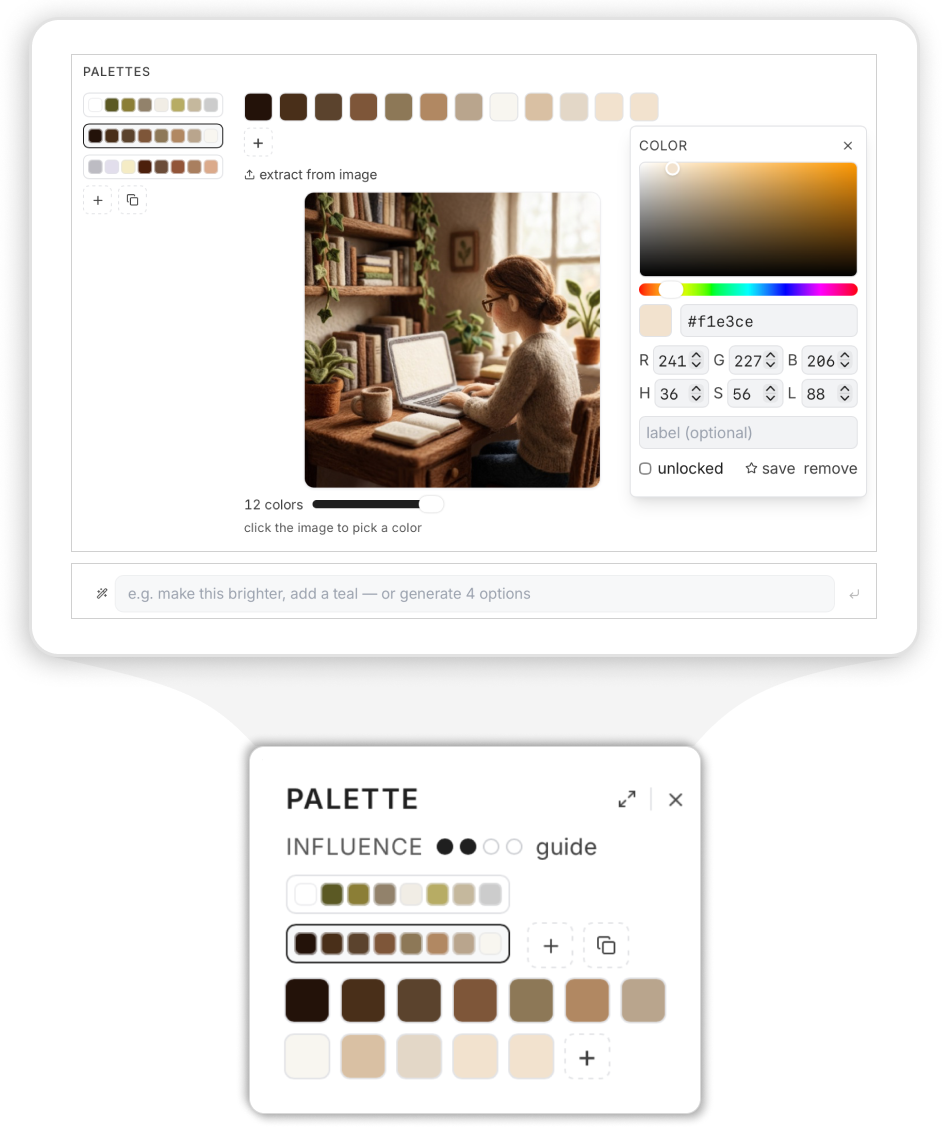}
    \caption{\texttt{Palette} node editor supports (1) direct manipulation, (2) extract from reference image, and (3) generate multiple options from prompt}
  \end{subfigure}\hfill
  \begin{subfigure}[t]{0.31\textwidth}
    \includegraphics[width=\linewidth]{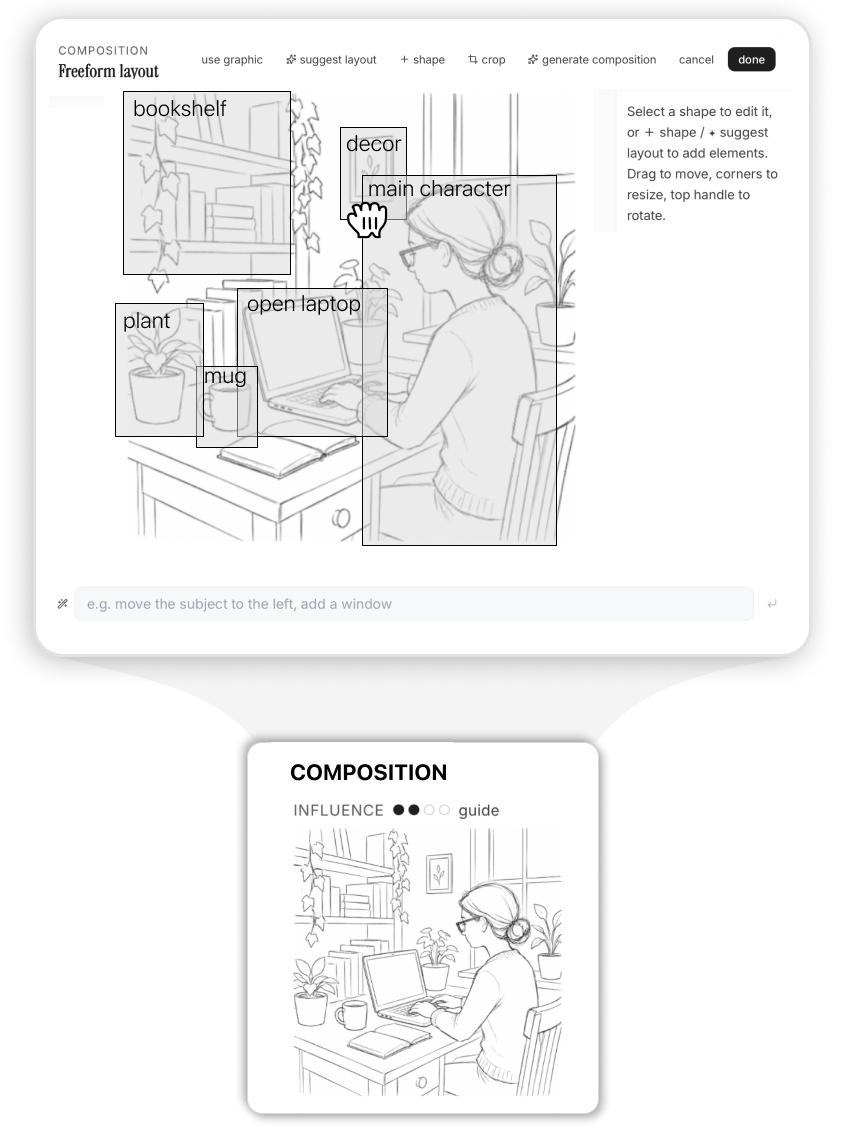}
    \caption{\texttt{composition} node's \texttt{scene} mode editor supports (1) freeform sketch, (2) direct manipulation (drag/resize) on objects, and (3) edit via prompt}
  \end{subfigure}\hfill
  \begin{subfigure}[t]{0.31\textwidth}
    \includegraphics[width=\linewidth]{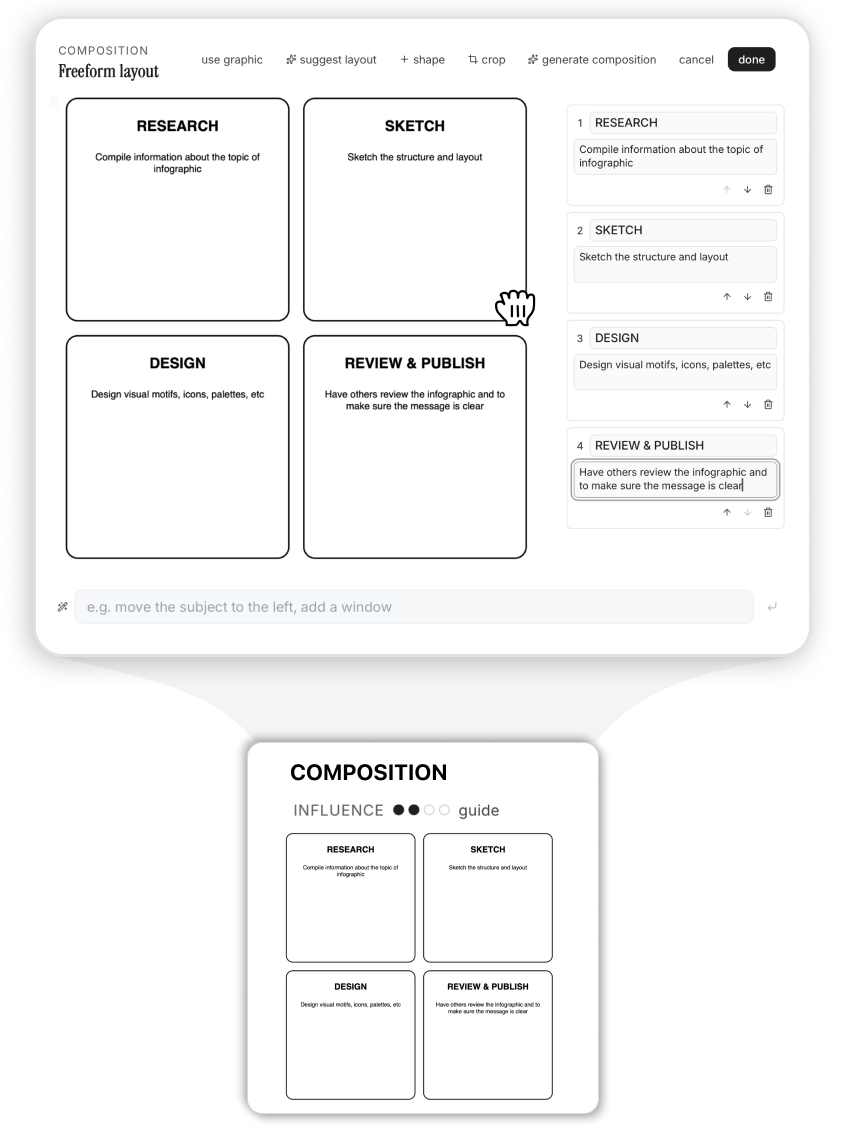}
    \caption{\texttt{composition} node's \texttt{graphic} mode editor allows users to (1) directly manipulate layout via wireframes and (2) edit the text directly in each field.}
  \end{subfigure}
  \caption{Each node in the workspace expands into a detailed editor workspace, supporting fine-grained, multi-modal interaction and manipulation of the visual element.}
  \Description{Examples of modality-specific plan-element editors. The palette editor supports direct color selection, extraction from a reference image, and generated palette options; the scene-composition editor supports sketching and spatial manipulation; the graphic-composition editor supports arranging wireframe regions and editing their text.}
  \label{fig:editor-collage}
\end{figure*}

The resulting plan is represented as a collection of plan elements (also referred to as nodes) where each node exposes an editor appropriate to the design decision. In contrast to the nodes in a typical node based user interface, plan element nodes are presented to the user as a set without connectors between the nodes. Each is meant to represent a single design decision and the system synthesizes them into a final image generation prompt. We chose this design to avoid the high degree of complexity associated with node-based graphical editing while presenting the user with individual, semantically meaningful controls over final output.  \autoref{tab:node-editors} outlines all the plan nodes currently present in the system. \autoref{fig:editor-collage} shows interaction details for the palette and composition nodes. By interacting with these nodes the user can control specific design decisions that impact the final image generation.

\paragraph{Plan element customization}
The relevant dimensions of a creative task may change as an idea develops, and users may want different degrees of structure at different moments (DG1). Accordingly, the set of plan elements is itself editable: users can add new plan elements, remove ones that are no longer useful, or simplify and elaborate the plan without reconstructing it. Additionally, users can also add ``reference'' nodes to bring in visual or textual assets that they already have in order to influence or include in the final generation. When a user adds a \texttt{reference} node, they also describe its purpose via a text description. For example in \autoref{fig:vis-branch}, the user uploads a \textit{reference} image for the data visualization icons, which are then reflected as the data markers in the final generated image.

\paragraph{Per-element influence dial.}
Creative exploration often contains decisions with different degrees of constraint: some are fixed constraints that must be followed, while some are exploratory directions (DG2). For example, a product designer working on a prototype mock-up may require the \texttt{subject} (product) to be represented accurately in the final image, while leaving \texttt{lighting} and \texttt{palette} for exploration.
In \sysname each node can be assigned one of four different levels of influence:
\begin{itemize}
    \item \texttt{None}: Excludes the node from final image generation without removing its visual representation from the UI.
    \item \texttt{Guide}: Treats the node as a loose direction that the model may reinterpret or adapt.
    \item \texttt{Strong}: Prioritizes close adherence to the node while allowing limited adaptation for visual coherence.
    \item \texttt{Locked}: Treats the node as a fixed constraint that should be preserved as faithfully as possible.
\end{itemize}
Together, these settings create a balance between allowing users to establish constraints while leaving other aspects open to interpretation, leveraging the model's generative capacity to propose possibilities that are not fully specified.
\begin{figure}[t]
    \centering
    \includegraphics[width=\linewidth]{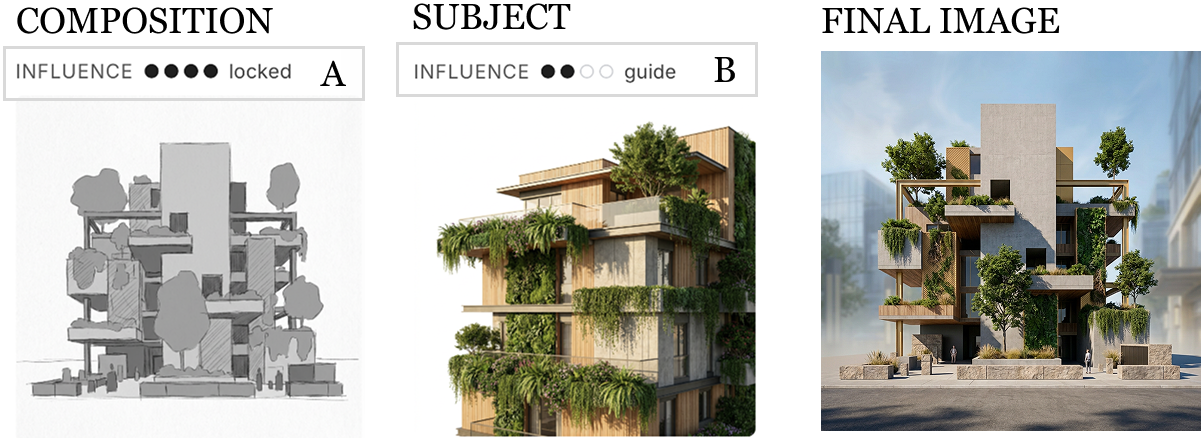}
    \caption{\textit{Composition} node (A), set to \texttt{locked}, is strictly followed in the final image, while \textit{subject} node (B), set to \texttt{guide}, is taken as reference but re-imagined.}
    \Description{Different influence levels applied to plan elements. A locked composition specifies the spatial arrangement of the scene, while a subject set to Guide provides a looser visual reference; the final image closely preserves the composition while reinterpreting the subject.}
    \label{fig:influence}
\end{figure}

\subsubsection{Prompt Synthesis and Final Image Generation}

When the user is satisfied with the plan, they click the ``Generate'' button to
produce the final image. \sysname\ compiles the plan elements into a single
structured, multimodal prompt for the image-generation model. The prompt opens
with a high-level description of the desired image together with global settings
such as the aspect ratio, followed by a textual representation of each plan
element, interleaved with the preview images generated for image-based elements.

How each element is serialized depends on its type. Text-only elements such as
\texttt{palette} and \texttt{typography} provide a description of the
element's role along with concrete details such as hex color codes or font
specifications. Image-based elements such as \texttt{subject}, \texttt{background},
and \texttt{composition} additionally attach the preview image generated in the
Plan Elements view, paired with instructions for how that image should be used in
the final generation. For example, a \texttt{background} node guides the
background rather than any foreground subject it may also depict. Where
applicable, these elements also contribute structured details, such as the
normalized bounding-box coordinates of a \texttt{composition}. Each element is
also serialized with its user-selected influence level, so that the prompt
reflects the relative influence the user assigned to it.

Once the prompt is constructed and sent to the model, the final image is rendered
in the ``Image'' component to the right of the Plan Elements. This component also
retains a history of generated images as the user edits the plan and generates
new images, supporting exploration and iteration.

A full example of this final prompt structure is included in Appendix \ref{app:final_prompt}.

\section{Examples}
\begin{figure*}[t]
    \centering
    \includegraphics[width=\linewidth]{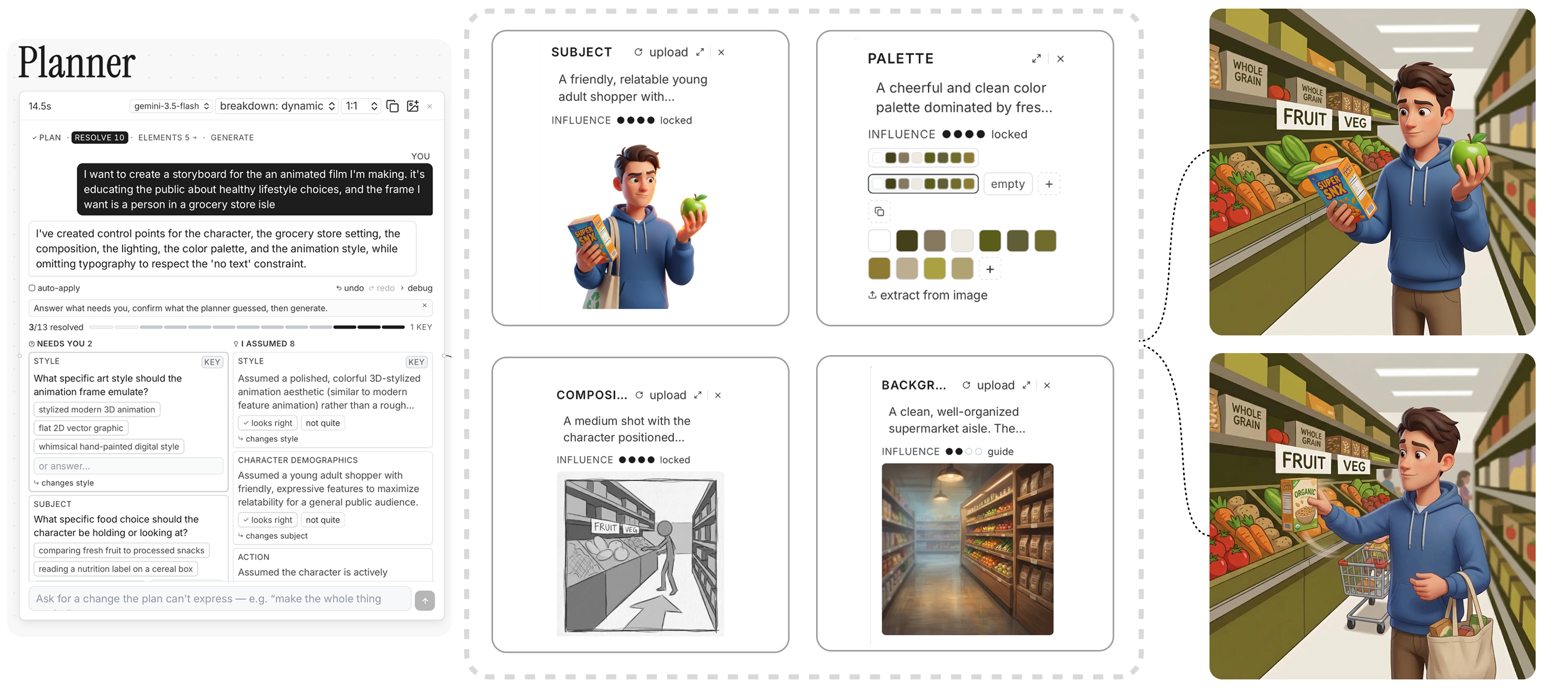}
    \caption{The \sysname planner surfaces assumptions about visual style and character demographics that the model is defaulting to, encouraging conscious reflection of these design choices. Once these are settled, the designer can create a storyboard by varying \texttt{composition} to generate each keyframe.}
    \Description{Storyboard generation after resolving model assumptions. The planning interface surfaces assumptions about character demographics and visual style; the resulting plan elements are then reused while composition is varied to produce two storyboard keyframes.}
    \label{fig:storyboard}
\end{figure*}

\sysname aims to provide a structured yet flexible and extensible interaction paradigm for a variety of visual tasks. Below, we showcase a selected set of example images generated using \sysname's workflow: stylized data visualization, a consistent set of architectural renderings, and an animation storyboard. In each of the use cases, by separating design aspects such as composition sketch, character, color palette, or typography, \sysname allows creators across these domains to ideate and iterate on these aspects independently, while progressing toward their goal with fine-grained control provided by \sysname. Additionally, in ~\autoref{fig:storyboard} we show how \sysname's planner helps the creator expand their conceptualization of a character by prompting the users to consider previously omitted design decisions.

\begin{figure*}[t]
    \centering
    \includegraphics[width=\linewidth]{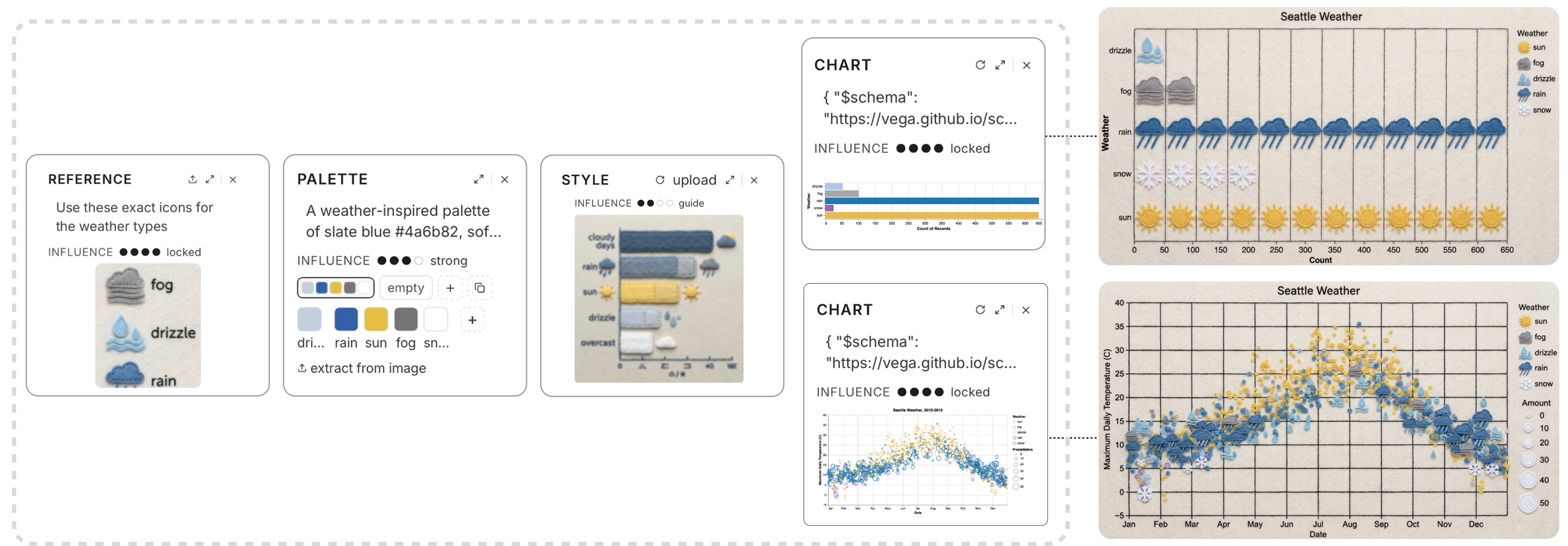}
    \caption{Creator makes different data visualizations rendered in a consistent, customized style. The creator explores palette and style, and provides fine-grained detail about the icon using the \texttt{reference} node. Then, she generates the two stylized visualizations by varying only the dataset in the \texttt{chart} node while holding other aspects consistent.}
    \Description{Consistent styling across two data visualizations. Reference, palette, style, and chart plan elements are shown alongside two generated charts that use different datasets while preserving a shared visual style, palette, and icon treatment.}
    \label{fig:vis-branch}
\end{figure*}

\paragraph{Stylized Data Visualization}
A data visualization designer uses \sysname to explore novel ways to stylize charts. In \sysname, the \texttt{planner} creates relevant dimensions for her task: \texttt{chart}, \texttt{style}, and \texttt{palette}. She manually adds a \texttt{reference} node to provide fine-grained visual specification for the icons, as well as annotate how her reference should be used. \autoref{fig:vis-branch} shows that by providing different datasets in the \texttt{chart} node but fixing the rest of the nodes, she creates consistently stylized renderings of the two datasets.

\begin{figure*}[t]
    \centering
    \includegraphics[width=\linewidth]{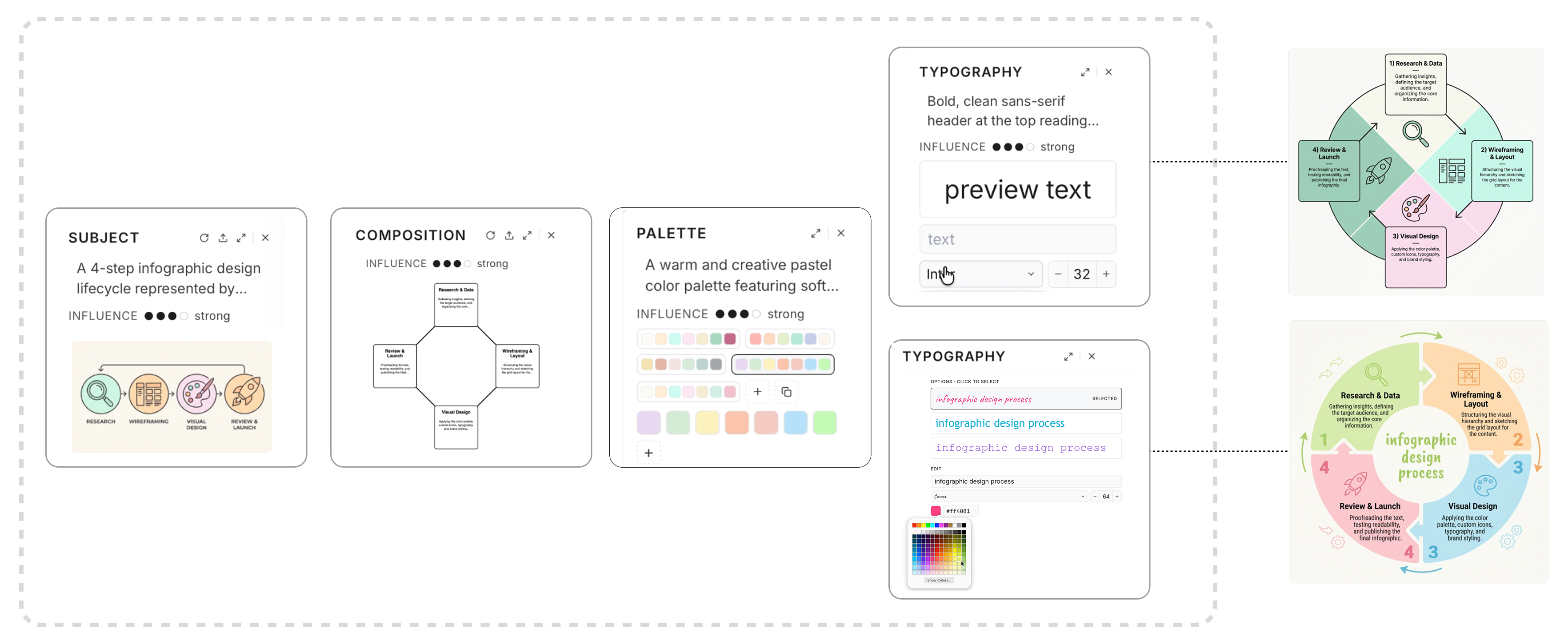}
    \caption{Infographic designer uses \sysname to brainstorm and ideate on different ways to illustrate her idea by specifying \texttt{subject} (icons), a circular \texttt{composition}, and a pastel \texttt{palette}, setting them as guidance to leave room for surprise, and varying different \texttt{typography}. \sysname generates two different designs that reflect different fonts while maintaining the layout, palette, and icons (right).}.
    \Description{Infographic variations produced by changing typography. Two generated infographics preserve the same icon subjects, circular composition, and pastel palette while using different font treatments.}
    \label{fig:infographic-example}
\end{figure*}

\paragraph{Infographic Design}
A designer uses \sysname to make an infographic about the infographic-design process.
After settling on decisions for \texttt{subject} (icons), a circular \texttt{composition}, and a pastel \texttt{palette}, she wants to experiment with different fonts. ~\autoref{fig:infographic-example} (top) shows the generated result using the default \texttt{typography} node. She then customizes this node by asking the agent to generate ``three playful font options" (\autoref{fig:infographic-example} bottom). Since she sets the influence level of composition to "\texttt{strong}" but not "\texttt{locked}", she also gets to explore slight variations in layout while maintaining the overall circular flow.

\begin{figure*}[t]
    \centering
\includegraphics[width=\linewidth]{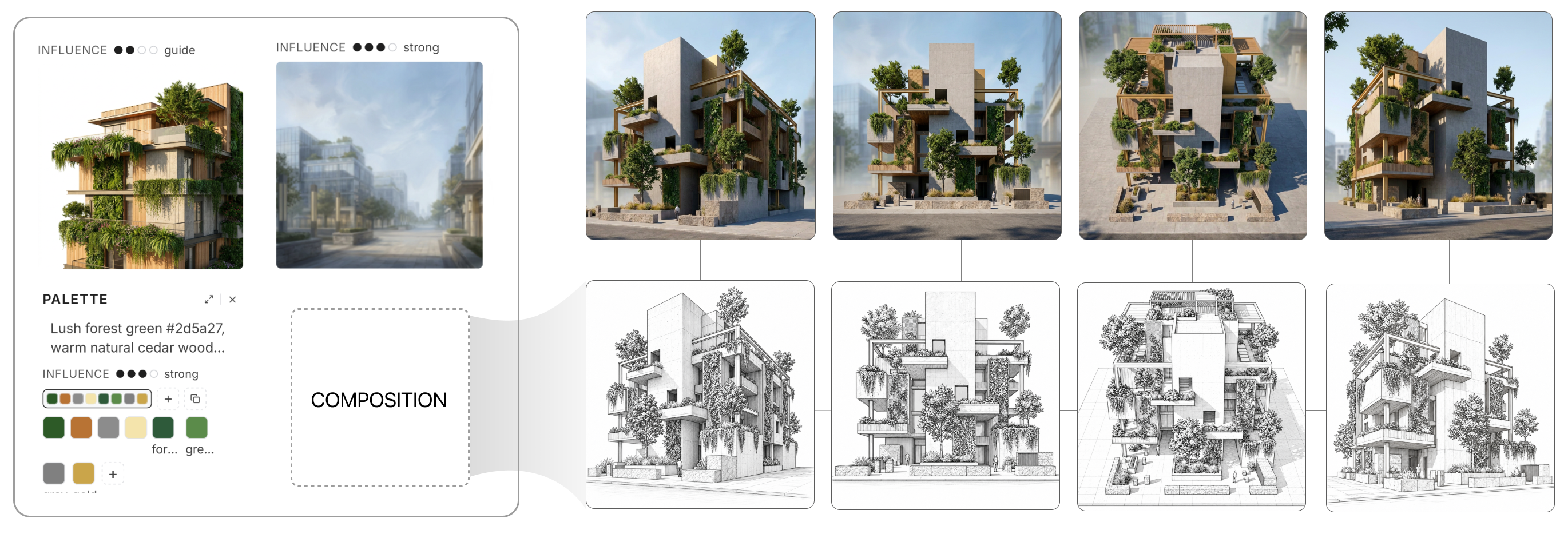}
    \caption{User creates a consistent set of architectural renderings from different angles by fixing the \texttt{background}, \texttt{subject}, \texttt{lighting}, while using a different \textit{composition} node for each.}
    \Description{Architectural concept rendered from multiple viewpoints. Shared subject, background, lighting, and palette plan elements are paired with different composition specifications to generate four views of the same plant-covered building.}
    \label{fig:architecture-variations}
\end{figure*}

\paragraph{Architecture Concept Rendering}
An architecture graduate student uses \sysname to create a set of visuals to communicate his vision for a research project on climate-conscious architecture. After he describes the project goal and his ideas, the \texttt{planner} prompts him to think about decisions he has not yet specified, including the site and region that this project is set in, the rendering style, and the background. \autoref{fig:architecture-variations} shows that after iterating on \texttt{subject} (building), \texttt{background}, and \texttt{palette}, he can generate multiple views by interacting with the \texttt{composition} node via natural language, describing ``isometric", ``side", ``elevated", and ``axonometric" views of the same building. \sysname generates a set of consistent renderings that follow the existing subject and background, while also following these view specifications.

\paragraph{Animation Storyboard}
A motion designer uses \sysname to create a storyboard to communicate an idea for a short animated film to her teammates. She first writes down the rough idea in the \texttt{planner}, which  prompts her to think about character demographics which she had not yet considered. The planner also surfaces a few assumptions about the illustration style (such as polished, 3D animation). Using these questions to reflect on her intention for the film, she iterates on each dimension and settles on the character and style suitable for their intended message.

%% file: sections/05_evaluation_and_results.tex
\section{User Study}
\label{sec:user-study}

To understand how \sysname supports creative workflows, we conducted a within-subjects interview study with \numparticipants participants who create imagery as part of their day-to-day work.
Participants included designers who create consumer-facing visuals, marketing professionals who design advertising materials, and engineers who create technical figures for scientific communication (\cref{tab:participants}).
We had each participant use \sysname and a chat-based interface to generate images for a task from their own work.
This allowed us to test whether \sysname supports real-world constraints of participants' creative practices.

\subsection{Study Design}

\paragraph{Participants}
\input{tables/participant-table}

To gain rich insights from experienced creators, we recruited \numparticipants participants who create imagery in a professional setting.
We identified relevant participants through calls within our organization and had them complete a pre-screening survey about their role, the type of imagery they create, and their experience with generative image tools.
To understand a range of perspectives, we filtered to participants who held varying titles (e.g., designers and engineers) and created diverse types of imagery (e.g., animations and marketing materials).
Participation in the user study was voluntary, and all participants gave informed consent.

\paragraph{Tasks}
Rather than imposing a predefined task, we asked participants to bring an image generation task from their own work.
This allowed us to evaluate \sysname in realistic scenarios with real-world constraints, because participants had a stake in the resulting image.
Participants brought a wide variety of tasks, including storyboards for animations, marketing graphics, images for presentations, and multi-panel technical diagrams.

\paragraph{Method}
We conducted each approximately 60 minute interview over video chat.
We began by discussing participants' current process for creating images and their experience with generative image tools.
Then, participants completed their task using a chat-based interface and \sysname, with condition order counterbalanced across participants.
We asked participants to think aloud during the study and had them reflect on their experience in both conditions.
With participant consent, we recorded screen and audio for each interview and collected interaction logs within each system.

\paragraph{Analysis}
To identify recurring patterns and insights from the study, we reviewed session recordings, think-aloud transcripts, and interaction logs.
First, we performed open coding to identify participants' experiences during each condition.
Then, we conducted an iterative affinity diagramming and thematic analysis to identify higher-level themes about how the two interfaces shaped participants' creative processes.
We used the results to structure \cref{sec:user-study-results}.

\subsection{Study Results}
\label{sec:user-study-results}

\subsubsection{Current generative image workflows do not match creators' needs.}
Today's generative image tools promise to handle the entire image creation process, turning an idea into a finished image the way a designer or artist would.
Yet despite participants' experience with generative image models and enthusiasm for their capabilities, none of our participants used them to create finalized images.
The 4/\numparticipants who used generative image models at all primarily relied on it for brainstorming, such as ideating new styles (P2) and generating mockups (P4).
While these AI-generated options helped participants see a breadth of ideas at once, participants did this infrequently and early in their workflows, before discarding the results and completing the rest of the design process manually.

The main barrier to adoption is that current generative image tools do not offer enough control or consistency to meet real-world constraints.
When creating visual assets, participants have strict requirements, such as using company-approved fonts and color palettes (P3) or matching a character style they own the intellectual property of (P2).
Since their images are often part of a larger design ecosystem, they also require consistency across designs, like sequential storyboard images (P3), icon families (P2), and slide deck figures (P7).
The need for control was consistent across all participants, from professional creators (P1, P2, P3, P4, P8) to business professionals (P5) and engineers (P7).

Current generative image tools lack the level of control needed for participants to meet these constraints and ensure cohesion.
As P4 described, generative image models can \pquote{get to 90\% really quickly and then dialing in that last 10\% is so difficult.}
As a result, while generating an image with current tools is easy, generating one that meets their design standards is nearly impossible.

\subsubsection{Planning supported participants' iterative design processes and prevented unwanted model assumptions.}

A commonly experienced disconnect (5/\numparticipants) between real-world image creation and generative image models is that most image generation tools assume that users have a fully specified design in mind at the outset.
Chat-based tools generate a fully rendered image after every prompt, regardless of whether the user has specified a few-word idea or a few hundred word description.
This contradicted participants' established design processes, which were highly iterative, starting from an intentionally vague idea that they refined over time as their vision became clearer.
\begin{quote}
    \pquote{I have a broad sense [that] I want something action focused... but I don't know specifically what I want, so I'm looking for help getting there and refining it along the way.}~--~P8
\end{quote}

As a result, chat-based interfaces often anchored participants to a particular image by making assumptions and filling in details on their behalf.
For example, while working in the chat interface, P5 prompted the model to depict a customer interaction using their workplace background.
Since P5 did not fully describe how the employee or customer should look, the model filled in their facial expressions, perceived demographics, and workplace uniforms.
Although these details influence how the image is perceived during employee training, seeing the model's choices anchored P5 to that design and they stopped making substantial iterations, saying \pquote{I didn't come into this with an explicit [idea],... so I guess at this point, I'm just reacting to what it's already created.}
These anchoring effects were so strong that even when participants had a particular design in mind, seeing the generated image could change their intended design without them noticing.
Before starting, P8 enthusiastically described wanting a clay style to match the playful tone of their existing marketing graphics.
But they forgot to include that instruction in their prompt to the chat interface, so the model generated an animated style instead.
Despite the image conflicting with their stated preference, P8 never corrected the style throughout the session.
When we pointed this out, they were surprised, saying \pquote{I, for some reason, throughout the conversation, did not ask it to change to [clay] style [...] let me try it now.}

In contrast, \sysname's planner gave participants control over their visual asset by making model assumptions explicit.
Before a participant could be anchored by a generated image, the planner surfaced underspecified aspects of the prompt, showed what the model would fill in by default, and let participants change it to match their expectations.
As a result, participants (P1, P2, P4, P5, P7, P8) used the planner to review and correct the model's implicit assumptions prior to generation.
These corrections ranged from simple preferences, like omitting text (P1) or using dark mode (P4), to critical design choices, like ensuring generated people matched their workplace demographics (P5).

The planner also expanded participants' thinking by surfacing important aspects of the design that they did not initially consider.
Participants were often surprised by the planner's questions about details they had not thought to include in their prompt, such as a character's clothing (P7) or the store lighting (P5).
Participants felt the planner \pquote{was asking very intelligent questions around what I wanted out of the image, which means I'm not wasting time and tokens generating an image [...] when we could be talking about what is expected and [...] our shared knowledge or lack there of} (P5).
Overall, the planning process was intuitive because it reflected participants' existing workflows, similar to \pquote{what you do in your head when you're starting to think of your drawing} (P2) and how they typically brainstorm with colleagues (P5, P8).

Interestingly, participants valued some amount of underspecification and did not want to specify every aspect of their design.
They had strong opinions on planner questions that related to their design taste or project constraints.
But when the planner surfaced decisions that participants did not feel strongly about, they were happy to skip the question and default to the model's assumption.
For example, when generating a slide deck figure, P7 decisively answered questions about the professional target audience, but left the model to determine the color palette, explaining \pquote{I'm not really good with colors, so that's the kind of thing I leave to the model to figure out.}

Even when participants allowed the model to fill in details on their behalf, surfacing the question in the planner was still valuable because it exposed design decisions that needed to be made.
Often seeing the model's assumptions helped participants clarify the design choice they actually wanted, and they would return to the planner to update the decision accordingly (P2, P4, P5, P8).
For instance, after seeing the model assume an isometric design, P4 realized they preferred a flat design and returned to the planner to correct it.

\subsubsection{\sysname's decomposition into plan elements matched the visual design process and made participants' interactions more intuitive.}
Our participants primarily interact with image generation models through chat interfaces that require them to describe visual properties in words.
This conflicts with their existing creative practices that center on visual communication, such as illustrating an idea (P4) or collaborating by editing a shared image (P3).
As a result, it was hard for participants to express their exact design requirements or make precise edits, with P3 describing \pquote{I want this very specific thing, [but] it's hard to explain in words.}
This was especially problematic for professional designers who found it frustrating \pquote{to spend time describing something when [they] could just make it} (P4).

In contrast, \sysname's plan elements broke the prompting process into visual design components that matched participants' mental models of image creation.
Since these elements draw on longstanding creative practices, thinking about an image in terms like subject, composition, and background felt intuitive.
\begin{quote}
    \pquote{The people who are best at using genAI are the people who can articulate their ideas at a level of detail that is extreme. So [for \sysname] to remove that as a requirement and make it visual makes it more accessible, because visual people are primed to think about it this way to begin with.}~--~P4
\end{quote}

By dynamically decomposing the plan into task-specific elements, \sysname fit a wide range of tasks and participant working styles.
For example, it automatically included a typography element for P4's technical diagram, a lighting element for P5's realistic scene, and four subject elements for P6's four-piece set of illustrations.
Participants also added and deleted elements to better fit their personal working style, such as adding a reference sketch (P3) or replacing a model-generated style with their own style example (P5).

Describing the image as a set of design elements also helped participants communicate with the model.
In particular, our design participants often work with other designers who they \pquote{can assume have a level of expertise already and know how to make [design] choices} (P4).
Whereas, generative image models lacked that shared expertise and often made poor decisions that professional designers never would.
Breaking the image into elements helped participants get more out of their AI collaborations, because \pquote{when you have been designing for a long time, so much of this that is second nature, ... so to break it down at this level reminds you, like oh I have to be really clear about everything} (P4).
Some participants found the decomposition so useful that they carried it into their written prompts, adding individual descriptions of the subject and composition (P1) to better communicate with the model.

Since the plan elements are inspired by two-dimensional creative practices like drawing and painting and our formative interviews with image experts, they fit most participants' use cases.
However, participants who work on three-dimensional videos and animations wanted additional elements to capture their three-dimensional vision, such as perspective, camera angle, and shot progression.
This validates our plan elements as core primitives of image design, while also suggesting future work incorporating primitives from video generation, such as by drawing on cinematography theory or letting AI models define novel element types when a user's task falls outside of the predefined set.

\subsubsection{Editing individual plan elements gave participants precise control over the generated image.}
Chat interfaces structure image generation as a linear conversation, where users submit an initial prompt and repeatedly request changes over multiple turns.
Because context compounds throughout the conversation, models often got stuck on a particular design and stopped making requested edits, forcing several participants to restart their chat session (P2, P5).
This linear workflow also lacked precise control because subsequent prompts reinterpreted the entire image, so minor requests like asking the model to \pquote{zoom in} could \pquote{completely change the layout} (P5).

Instead of a linear workflow, \sysname's plan elements acted as a visual toolkit that composed into a final image.
Rather than packing every idea into one prompt, participants worked through each element to make relevant design decisions, such as choosing a color scheme in the palette element (P2, P3) or working through layouts in the composition element (P1, P4).
Since previous design decisions were not implicitly included in the model's context, participants could revert unwanted decisions without the model getting stuck on a particular design.
As a result, participants kept editing and refining plan elements as they progressed towards a finalized image in \sysname, making 2.4x more edits to the model's generations ($p < 0.05$\footnote{$p$-value computed using a paired t-test.}) using \sysname ($\mu = 11.4$) than in the chat interface ($\mu = 4.8$).
\begin{quote}
    \pquote{It's a design process thing. [In] design you are constantly making something and then changing it. Editing, iterating, and getting feedback [are] such innate parts of the process. With a palette of options like this, it's so inviting to be able to tweak something and see how it changes.}~--~P4
\end{quote}

By mapping each plan element (e.g., composition) to a corresponding direct manipulation interface (e.g., draggable objects), \sysname allowed participants to make targeted edits.
For instance, P2 felt they had more control over their image's color when working in the palette element, saying \pquote{I really liked the color palette picker, because if I was copy pasting HEX codes into [a chat interface], it's just text, so I'm going to have no idea which one corresponds to green or brown.}
The influence dial on each element gave participants further control over the final image (7/\numparticipants adjusted it), such as letting P3 enforce the style of their reference image while treating their suggested palette as just a guide.

Being able to break an image into independently controllable elements also excited participants about using \sysname to facilitate design collaboration.
P3 and P4 act as consultants who create and redesign content for stakeholders, like scientists and corporate partners.
By locking certain elements (e.g., color, style) and leaving others editable (e.g., subject, composition), participants hypothesized that \sysname could \pquote{give agency to stakeholders to be more independent [...] [while] creating things under certain guidelines we set} (P3).

\subsubsection{By increasing participants' control over image generation, \sysname changed their relationship with AI}
\sysname increased participants' control over their generated image by surfacing underspecification and model assumptions via the planner, highlighting visually salient elements in the plan, and providing direct manipulation editors for each plan element.
As a result, 6/\numparticipants expressed wanting to use \sysname in their professional workflows to generate visual assets.
They wanted tools with \pquote{the fine grained control [\sysname] had, because [existing interfaces] feel like doing surgery with boxing gloves on} (P5).

Making precise edits that were incorporated in the generated image inspired participants to reflect on their AI collaborations more broadly.
After using \sysname, P7 reflected that they had grown complacent about ceding control to AI models, saying \pquote{I feel like maybe I've defaulted to just letting the model generate [images] automatically for me because I had no control.}
While they felt chat interfaces limited their creativity, \sysname let them express themselves and have more agency in human-AI co-creation.
\begin{quote}
    \pquote{Creativity is limited by the tools [you] have. [...] In a chat interface, the model takes your idea, shaves off all of its personality, [and] generates a very generic image, [whereas] with [\sysname] you are giving people the power to express themselves and putting the model back in its place.}~--~P7
\end{quote}

\input{sections/05_01_automated_evals}

%% file: tables/participant-table.tex
\begin{table}
\centering
\caption{
We evaluate \sysname with \numparticipants participants who create images as part of their day-to-day work.
Participants represent diverse roles and creative tasks that span design, engineering, and business.}
\Description{Roles and image-generation tasks of the eight study participants. Participants include graphic, UI/UX, and visual designers, an art director, an account manager, a machine learning engineer, and a marketing manager, with tasks ranging from branding and storyboarding to technical diagrams and marketing graphics.}
\label{tab:participants}
\small
\begin{tabularx}{\linewidth}{@{}rl>{\raggedright\arraybackslash}X@{}}

\addlinespace[2pt]
\textbf{{ID}} & \textbf{{Title}} & \textbf{{Task}} \\
P1 & Graphic Designer & Branding image of company products \\
P2 & UI/UX Designer & Video game stickers for a messaging app \\
P3 & Graphic Designer & Style frames for animation storyboarding \\
P4 & Visual Designer & Technical diagram for a presentation \\
P5 & Account Manager & Image for an employee training presentation \\
P6 & Art Director & Set of sales illustrations \\
P7 & ML Engineer & Teaser figure for a presentation \\
P8 & Marketing Manager & Marketing graphic for a mobile game \\
\end{tabularx}
\end{table}

%% file: sections/05_01_automated_evals.tex
\section{Pipeline Validation}

Given the significant transformation \sysname makes to user prompts before final generation we decided to validate that our pipeline does not significantly harm the capabilities of the underlying model. To do this we turn to automated benchmarks typically used in text-to-image (T2I) generation evaluation. While this is not an exact fit to the scenarios \sysname is designed for, i.e. \sysname is designed for a heavily interactive and iterative process, while automated benchmarks remove human involvement,  we nevertheless find these to be a useful diagnostic of how the underlying pipeline affects the behavior of the model.

We evaluate on two benchmarks from the T2I literature. \genaibench \cite{Li_2024_CVPR_genaibench} is a well established benchmark consisting of 1{,}600 real-world text prompts sourced from professional designers that cover ``essential visio-linguistic compositional reasoning skills such as comparison, counting and logic''. This benchmark comes with a metric, VQAScore, and an LLM evaluator to measure how accurately the generated image depicts the prompt. \qwenimagebench \cite{li2026qwen_image_bench} is a newer and more complex benchmark consisting of 1{,}000 prompts organized into 5 top level pillars (aesthetics, alignment, creative generation, quality and real-world fidelity) that are further decomposed into 56 evaluation facets. The \qwenimagebench taxonomy was co-developed with professional artists and designers who then worked with LLM assistance to create prompts that reflect realistic text-to-image generation goals. \qwenimagebench provides an LLM-as-judge evaluator (Q-Judger) that evaluates each generated image on up to 56 facets, with a rubric to produce either a N/A, fail, pass, excellent ratings. The mean number of evaluation facets per prompt-image pair in our evaluations is 24.87.

We compare benchmark performance under two conditions: \raw, where the prompt is sent directly to the model, and \plan where the prompt is sent to a ``headless'' version of reComposer that generates a result with no human intervention, relying fully on the decisions reComposer makes. We use \textbf{Gemini 3 Pro Image} for final image generation in both conditions and \textbf{Gemini 3.5 Flash} and \textbf{Gemini 3.1 Flash Image} for planning and intermediate image generation respectively in the \plan condition. While this ``headless'' mode is not the mode in which the system was designed to be used, we believe it does offer some insight into the performance of the pipeline underlying the system at scale. We grade image responses for each benchmark using the scorers and methodology provided by the benchmark authors.

\begin{table}[t]
\centering
\small
\setlength{\tabcolsep}{5pt}
\caption{Top-level results: \plan vs.\ \raw. Scores are VQAScore (0--1) for
  GenAI-Bench and the composite (0--100) for Qwen-Image-Bench. $\Delta =$ \plan
  $-$ \raw (paired, 95\% bootstrap CI).}
\Description{Top-level benchmark results for raw prompting and the reComposer planning pipeline. GenAI-Bench scores are 0.883 for raw and 0.884 for plan, a difference of +0.001 with a 95-percent confidence interval of -0.015 to +0.016. Qwen-Image-Bench scores are 60.2 for raw and 56.9 for plan, a difference of -3.4 with a 95-percent confidence interval of -4.1 to -2.6.}
\label{tab:toplevel}
\begin{tabular}{lrrc}
    \toprule
    Benchmark & Raw & Plan & $\Delta$ [95\% CI] \\
    \midrule
    GenAI-Bench      & .883 & .884 & $+0.001\ [-0.015,\,+0.016]$ \\
    Qwen-Image-Bench & 60.2 & 56.9 & $-3.4\ [-4.1,\,-2.6]$ \\
    \bottomrule
  \end{tabular}
\end{table}

\begin{figure}[t]
    \centering
    \includegraphics[width=\linewidth]{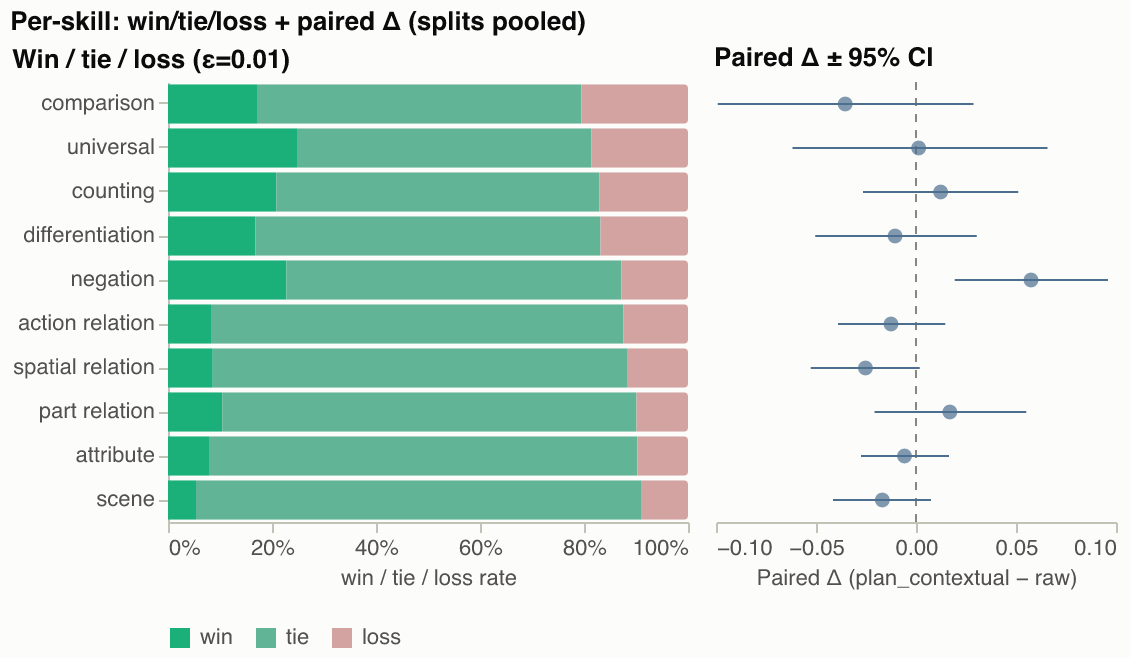}
    \caption{\genaibench performance by skill.}
    \Description{GenAI-Bench performance by compositional skill. For each of ten skills, horizontal stacked bars show plan-condition wins, ties, and losses relative to raw prompting, while adjacent points with 95-percent confidence intervals show paired score differences. Most differences are near zero, with negation showing the clearest advantage for the plan condition.}
    \label{fig:genaibench_skills}
\end{figure}

On \genaibench, the \sysname pipeline is statistically indistinguishable from raw generation. \autoref{fig:genaibench_skills} shows performance across the 10 skills this benchmark tests. On \qwenimagebench we see a regression of -3.4pp in benchmark scores in the plan condition compared to raw. (\autoref{tab:toplevel}).

\autoref{tab:qwen_facets} shows evaluation facets with statistically significant differences in the two conditions and reveals that a) a number of these are closely related and have to do with text rendering or placement and b) because fail/pass/excellent is mapped to 0/60/100 in the final scoring, failures are more heavily penalized than great results rewarded and the greater number of outright failures in the \plan condition does show large absolute differences in performance. We do also note that the judge fairly often uses different evaluation facets on the images for each prompt which may have some effect in the aggregation.

\begin{table}[t]
\centering
\small
\setlength{\tabcolsep}{3.5pt}
\renewcommand{\arraystretch}{0.95}
\caption{Per-facet verdict counts (fail/pass/excellent) for the 21 facets with differences between conditions. $\Delta n = n_{\text{plan}} - n_{\text{raw}}$ is the
difference in number of samples for this facet in each condition (+ means more prevalent in the plan condition).}
\Description{Qwen-Image-Bench results for 21 facets with significant differences between conditions. Columns report the score difference, difference in number of evaluated samples, and fail, pass, and excellent counts for raw and plan conditions. The largest decreases for the plan condition are in cross-lingual generation, text accuracy, information visualization, font, animals, and graphic design.}
\label{tab:qwen_facets}
\begin{tabular}{@{}p{2.5cm} r r c c@{}}
\toprule
Facet & $\Delta$ & $\Delta n$ & raw & plan \\
\midrule
Cross-lingual Generation  & $-41.7$ & $+32$ & 6/24/22    & 53/18/13   \\
Text Accuracy             & $-23.9$ & $-58$ & 81/54/220  & 138/47/112 \\
Information Visualization & $-23.5$ & $-7$  & 21/22/58   & 43/17/34   \\
Font                      & $-15.6$ & $-17$ & 6/45/32    & 13/41/12   \\
Animals                   & $-15.5$ & $-2$  & 58/9/43    & 74/8/26    \\
Graphic Design            & $-13.2$ & $-5$  & 3/48/80    & 14/60/52   \\
Objects                   & $-11.9$ & $-1$  & 165/97/104 & 211/89/65  \\
Text Layout               & $-10.5$ & $-66$ & 11/165/165 & 29/155/91  \\
Anatomical Fidelity       & $-8.2$  & $-31$ & 380/69/56  & 403/46/25  \\
Style Control             & $-8.0$  & $+9$  & 57/289/208 & 99/304/160 \\
Composition Relationship  & $-8.0$  & $+1$  & 161/549/226& 235/553/149\\
Logical Resolution        & $-6.7$  & $-9$  & 87/99/52   & 94/108/27  \\
Naturalness               & $-6.5$  & $-5$  & 324/447/166& 385/440/107\\
Contact Interaction       & $-5.7$  & $-10$ & 182/268/22 & 218/228/16 \\
2D Space                  & $-4.6$  & $-14$ & 44/552/86  & 78/531/59  \\
Imagination               & $-3.4$  & $+8$  & 6/136/52   & 11/147/44  \\
Size                      & $-2.9$  & $-27$ & 23/475/24  & 37/448/10  \\
Physical Logic            & $-2.9$  & $-10$ & 605/259/65 & 634/235/50 \\
3D Space                  & $-2.4$  & $-9$  & 82/627/73  & 106/603/64 \\
Material Texture          & $-2.2$  & $-15$ & 43/466/407 & 67/447/387 \\
Resolution                & $-0.6$  & $0$   & 1/899/38   & 1/913/24   \\
\bottomrule
\end{tabular}
\end{table}

\begin{figure}[t]
    \centering
    \includegraphics[width=0.85\linewidth]{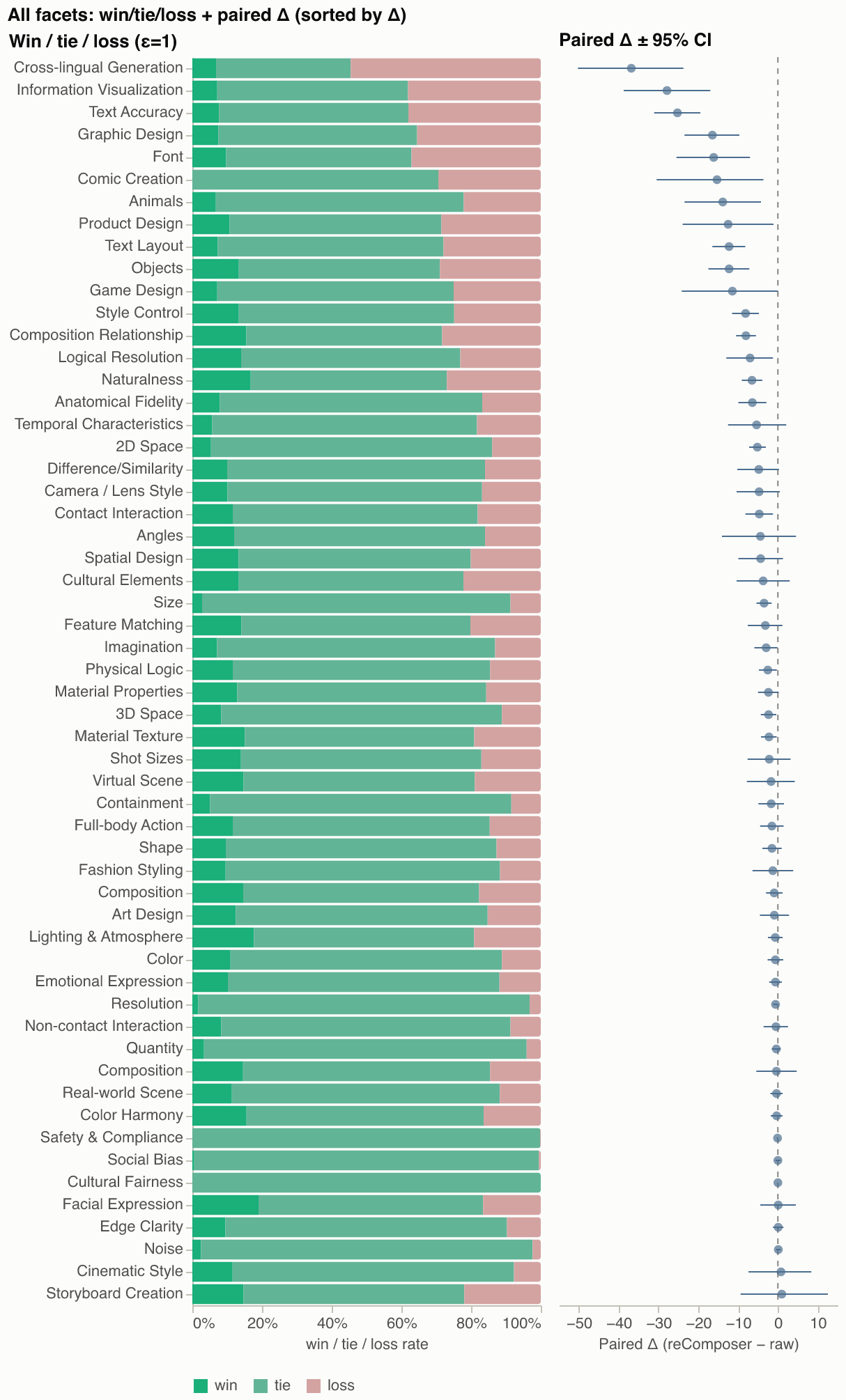}
    \caption{\qwenimagebench win/loss/tie rates for paired judgments. Only one facet has a loss rate more than 50\%}
    \Description{Qwen-Image-Bench paired comparison by evaluation facet. Horizontal stacked bars show plan-condition wins, ties, and losses relative to raw prompting, with adjacent points and 95-percent confidence intervals showing paired score differences. Ties dominate most facets, and only one facet has a loss rate above 50 percent.}
    \label{fig:qwen_image_facets}
\end{figure}

If we look at paired judgments (samples where the evaluation facet is triggered in both conditions) we can compute per-facet win/tie/loss rate of the plan condition compared to the raw condition. This is shown in \autoref{fig:qwen_image_facets} (right) where we observe that only one category has a loss rate of greater than 50\% and the plan condition otherwise being preferred or tying.

\begin{figure}[t]
    \centering
    \includegraphics[width=\linewidth]{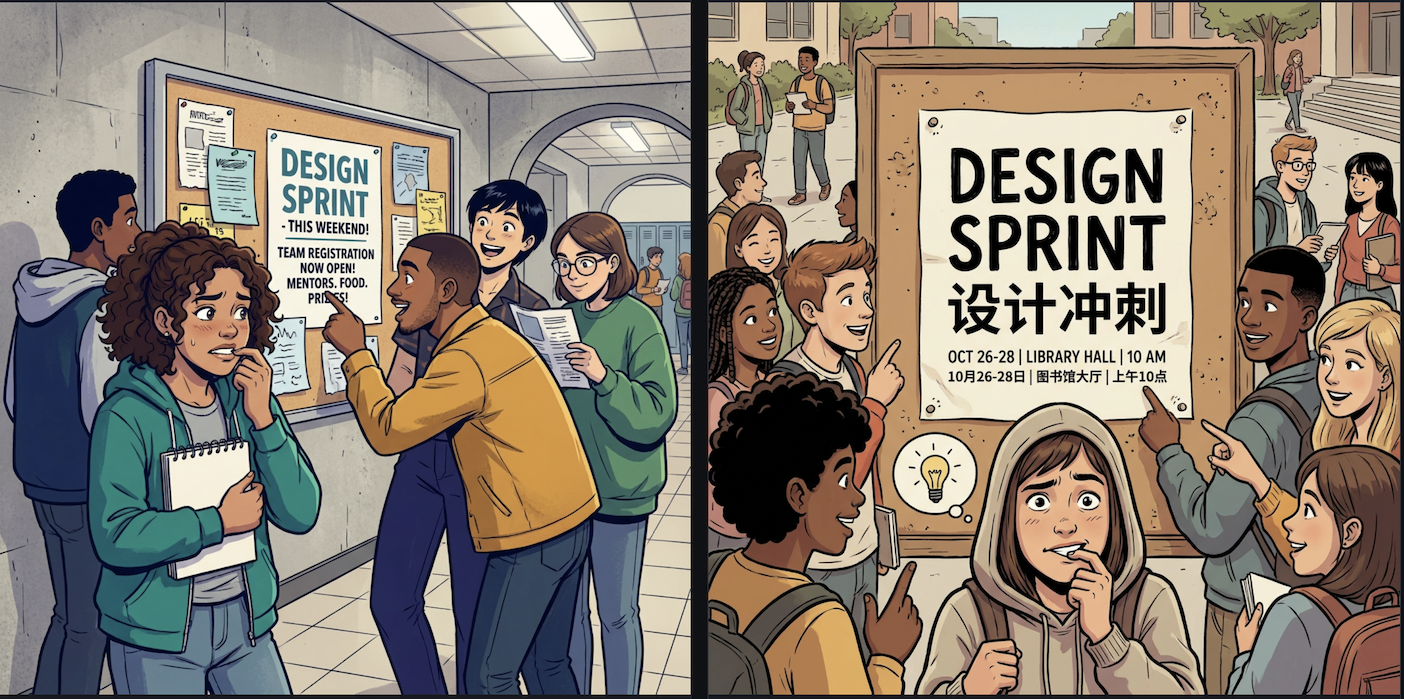}
    \caption{Plan condition on left, raw on right for the prompt \textit{``Comic scene: students gather at a campus bulletin board reading a poster titled “DESIGN SPRINT  \cjkimg{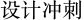}” with bilingual date/location and visible whitespace. One student bites their lip in nervous excitement with subtle expression. Require legible accurate poster text, realistic wrinkled paper and pushpin stress, diverse non-repetitive faces, and youthful excitement.''} The \plan condition image fails to include the requested Chinese text on the message board and is scored lower. However we note that the overall composition of this image is more realistic.}
    \Description{Failure case for bilingual text generation. The reComposer plan output and raw-prompt output both depict students around a campus bulletin board; the plan output omits the requested Chinese text, while the raw output includes the bilingual 'DESIGN SPRINT' poster more accurately.}
    \label{fig:qib_text_1}
\end{figure}

\begin{figure}[t]
    \centering
    \includegraphics[width=\linewidth]{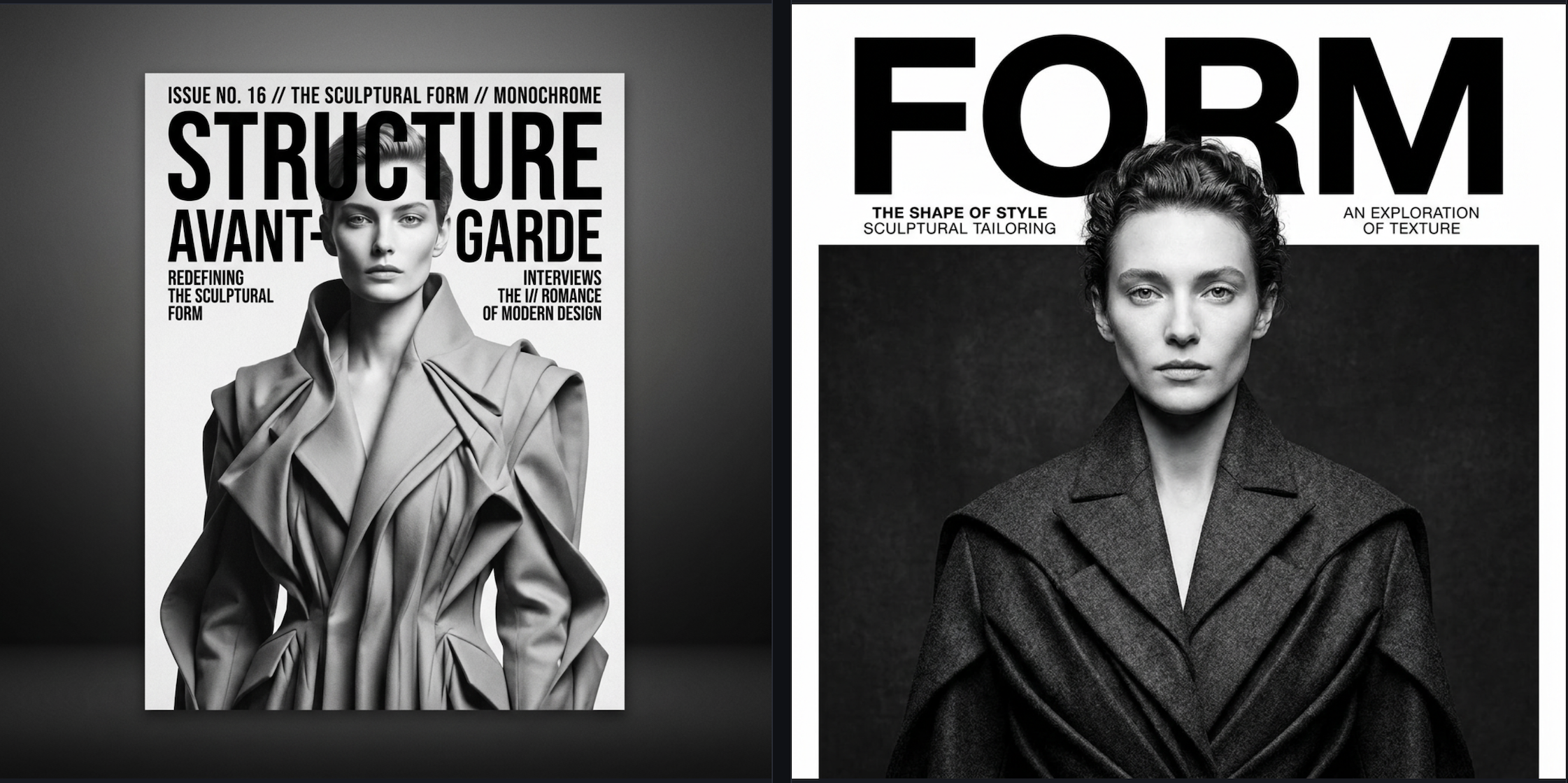}
    \caption{\plan condition on left, \raw on right for the prompt \textit{``B/W magazine cover: monochrome portrait, person in sculptural overcoat with calm gaze; big title “FORM” and multi-line small subheads. Require high face/hair quality; sharp professional typography; strong contrast without losing detail.''} The \plan condition includes the word form but only as part of a subtitle and adds extra text (albeit producing a more typical magazine cover layout).}
    \Description{Failure case for magazine-cover typography. The reComposer plan output uses 'FORM' as secondary text and introduces additional unintended text, while the raw-prompt output more accurately renders 'FORM' as the large requested title.}
    \label{fig:qib_text_2}
\end{figure}

\begin{figure}[t]
    \centering
    \includegraphics[width=\linewidth]{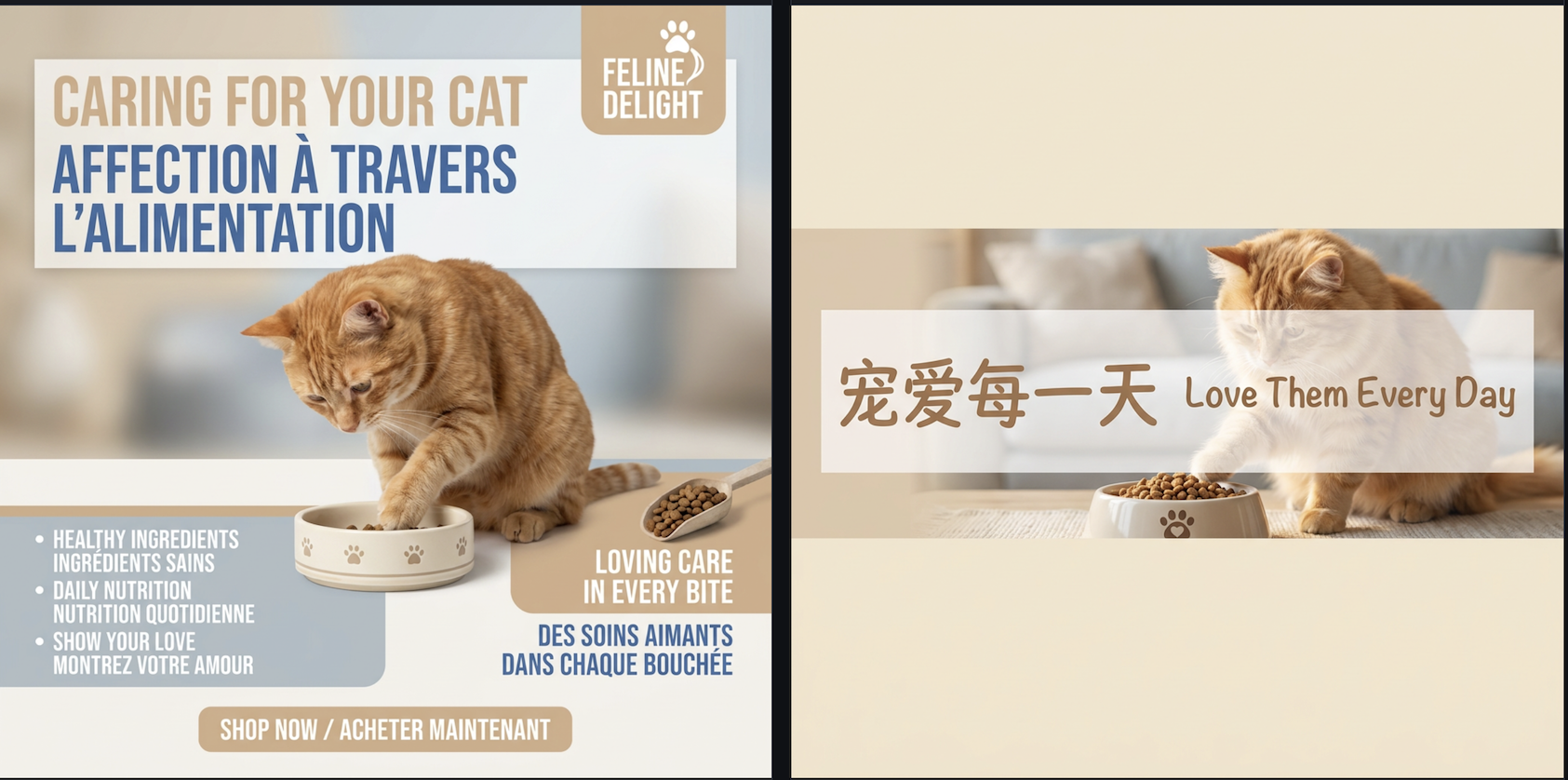}
    \caption{\plan condition on left, \raw on right for the prompt ``Generate a pet product e-commerce banner: left—“\cjkimg{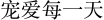}” in soft rounded font with slightly hand-drawn curves; right—“Love Them Every Day” matching weight and line height, bilingual baselines strictly aligned. Background: an orange cat gently pawing a branded cat food bowl. Text area overlaid with 15\% opacity white mask for high readability. Color palette: warm beige and light gray-blue.'' Then \plan condition result uses the wrong pair of languages for the bilingual text.}
    \Description{Failure case for bilingual banner text. Both outputs depict an orange cat in a pet-product advertisement, but the reComposer plan output uses an incorrect pair of languages for the requested bilingual text.}
    \label{fig:qib_text_3}
\end{figure}

Manual inspection of failures in these worse performing facets shows that they are often caused by a failure to render the exact text provided in the prompt in the final image, and this causes multiple related evaluation facets to be scored low. Placeholder text from the preview images from various nodes in the plan elements is often rendered in the final image either instead of or in addition to the requested text and is penalized. And background text is often rendered at a low resolution (see Figures \ref{fig:qib_text_1} \ref{fig:qib_text_2} \ref{fig:qib_text_3} for examples). We note that in the interactive workflow reComposer is designed for, these major textual elements are typically added or edited by the user in specific plan-element nodes, such as typography or composition, once the initial design direction is set. So while these results do indicate room for improvement, since \sysname is not designed for non-interactive generation we only consider these results as diagnostic pointers that represent the floor on the behavior of the system.

Overall we observe that even in this non-interactive setting, the pipeline generally produces images comparable to that of the underlying model.

%% file: sections/06_discussion_and_conclusion.tex
\section{Discussion, Limitations and Future Work}
Our user study findings provide promising evidence of our interaction paradigm, and highlight opportunities and challenges for treating intermediate creative decisions as editable representations in generative workflows.

\paragraph{Going beyond prompt-based interaction is important for advanced creative work}

Our findings reinforce prior work showing that creating with generative models benefits from interaction beyond only natural-language prompting~\cite{shi2025brickifyenablingexpressivedesign}. \sysname's user study findings suggest that the opportunity is not simply to expose more control, but designing these controls around how creators think about their process. \sysname dynamically surfaces relevant semantic decisions in the form of plan element nodes, and allows the user to add, remove and edit them. As a result, the interface evolves with the users' creative process, rather than requiring the user to conform to a static structure for specifying intent.

\paragraph{Editable Planning provides a shared representation for Human-AI Co-Creation}
\label{discussion:intent}
\sysname's user study findings demonstrate how an intermediate plan can serve as a shared representation that grounds human-AI interaction. Users directly manipulated individual design decisions, while the planning agent operated over the same representation when broader changes are easier to express conversationally. For example, a user may know that an image should feel 'softer' without fully knowing whether this should manifest as changes to palette, lighting, composition, or style. Rather than directly regenerating the image from this instruction, the agent can interpret the user's partially specified intent and propose coordinated changes to relevant plan elements, keeping the traces visible and revisable. This suggests a different affordance for dialogue in generative interfaces: instead of a modality for users to optimize a prompt toward an output, \sysname demonstrates how it can be a modality to help users translate high-level intentions into explicit, manipulable decisions.

\paragraph{Users desire both controllability and complementarity with generative systems}
Our user study results point to an overarching tension between control and complementarity in human-AI collaborative systems.
For example, to increase user control over image generation, we explicitly designed \sysname's planner to surface prompt underspecification and the model's implicit assumptions.
While we could have required participants to answer every planner question so the model did not make any assumptions about their intent, this would have forced participants to make design decisions before they were ready, prevented them from leveraging the model's decisions to complement their thinking, and ultimately limited their creativity.
Instead, we aimed to increase user agency by letting participants choose which decisions to make themselves, which to leave to the model, and when to change their mind.
\begin{quote}
    \pquote{[I want to] wield [AI] tools in a way that is still very human driven, [where] we use the algorithmic randomness in a way that is compelling and interesting as opposed to letting the technology drive on its own.}~--~P4
\end{quote}

\paragraph{Generative AI for creators should provide access to process in addition to high quality output}
\label{discussion:process}
One participant, after using \sysname, questioned whether they had previously let the AI take over too much before, they reflected that their engagement with the process before was limited because they knew they could only do so much in a prompt based system, but seeing new tools like \sysname in turn changed how they will engage with the process (P7). These comments suggest that making intermediate decisions visible can also make the division of contribution between user and model more visible. Prompt-to-output interfaces can obscure many of the decisions delegated between an initial prompt and a finished output. By externalizing some of these decisions, generative interfaces may create opportunities for users to consider, revise, or intentionally delegate them. This raises a broader design question of not only which parts of creative work can be automated, but which parts are valuable for people to continue engaging with.

\paragraph{User study limitations}
Our study participants were primarily designers and creators recruited from a U.S.-based technology company. While participants brought experience from different creative roles and practices, other creators from different domains may differ in how they conceptualize their process, which intermediate decisions they exposed, and how much structure or model initiative they prefer. Future studies could validate \sysname with a broader range of domain creators, and extend the set of controls for broader support.

Additionally, participants interacted with \sysname in a first-use study and for a limited period of time. Many participants expressed interest in spending more time exploring the \pquote{full potential of the tool} (E3), suggesting that the current findings may not capture the full set of interactions and experiences users would have with \sysname over a longer period of time. A longitudinal study examining how \sysname integrates into existing creative workflows could provide valuable insights.

\paragraph{Extending Editable Planning Beyond Visual Creation}
\label{discussion:generalize}
Although \sysname instantiates editable planning in visual asset creation, the underlying interaction paradigm is not necessarily constrained to visual generation. Increasingly, people collaborate with generative AI systems on tasks that unfold across multiple intermediate decisions and reasoning spaces that demand more than a single request and response. Writing, programming, and scientific workflows similarly require high-level goals to be translated into a changing set of intermediate decisions, many of which may otherwise remain implicit in prompts or be made by the model. Editable planning suggests an alternative to both extremes of requiring users to specify every step themselves or delegating the entire process to an AI system. Future interfaces might explore how we can foster synergetic human-AI collaboration by selectively externalizing decisions where human judgment, reflection, and ownership matter, while allowing others to be delegated.

\section{Conclusion}
We present \sysname, an image generation tool that operationalizes editable planning as a process-oriented design paradigm for controllable image generation. Grounded in real-world principles and workflows of visual designers and image generation practitioners, \sysname exposes semantic decisions throughout the generative process, enabling users to engage with and shape the intermediate creative choices rather than delegating the entire process to the model. In a user study with \numparticipants participants, our findings suggest that intermediate plans and visual artifacts can provide meaningful control and transparency, while fostering more reflective and process-aligned creative practice in human-AI co-creation.

%% file: sections/21_appendix_b.tex
\section{Generated Outputs}
\subsection{Samples from Gemini chat iterations}
\begin{figure}[H]
    \centering
    \begin{subfigure}{\columnwidth}
    \includegraphics[width=0.6\columnwidth]{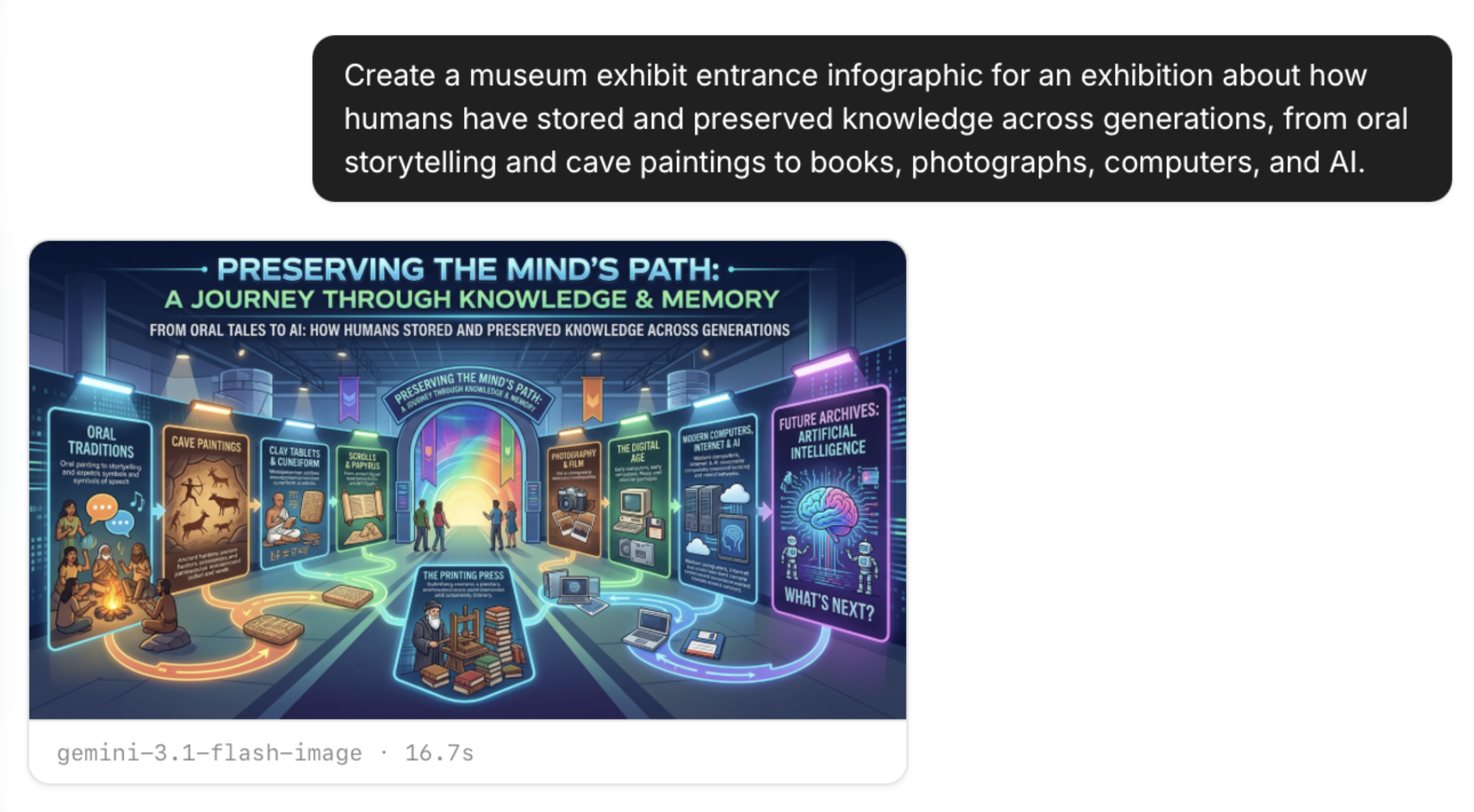}
        \subcaption{Initial prompt. The first sample fixes the image's composition -- an isometric corridor lined with exhibit panels.}
        \label{fig:gemini-initial}
        \Description{A chat prompt asking for a museum exhibit entrance infographic, and the returned image: an isometric view down a corridor
        lined with illuminated exhibit panels.}
    \end{subfigure}
    \\[1ex]
    \begin{subfigure}{\columnwidth}
        \includegraphics[width=0.6\columnwidth]{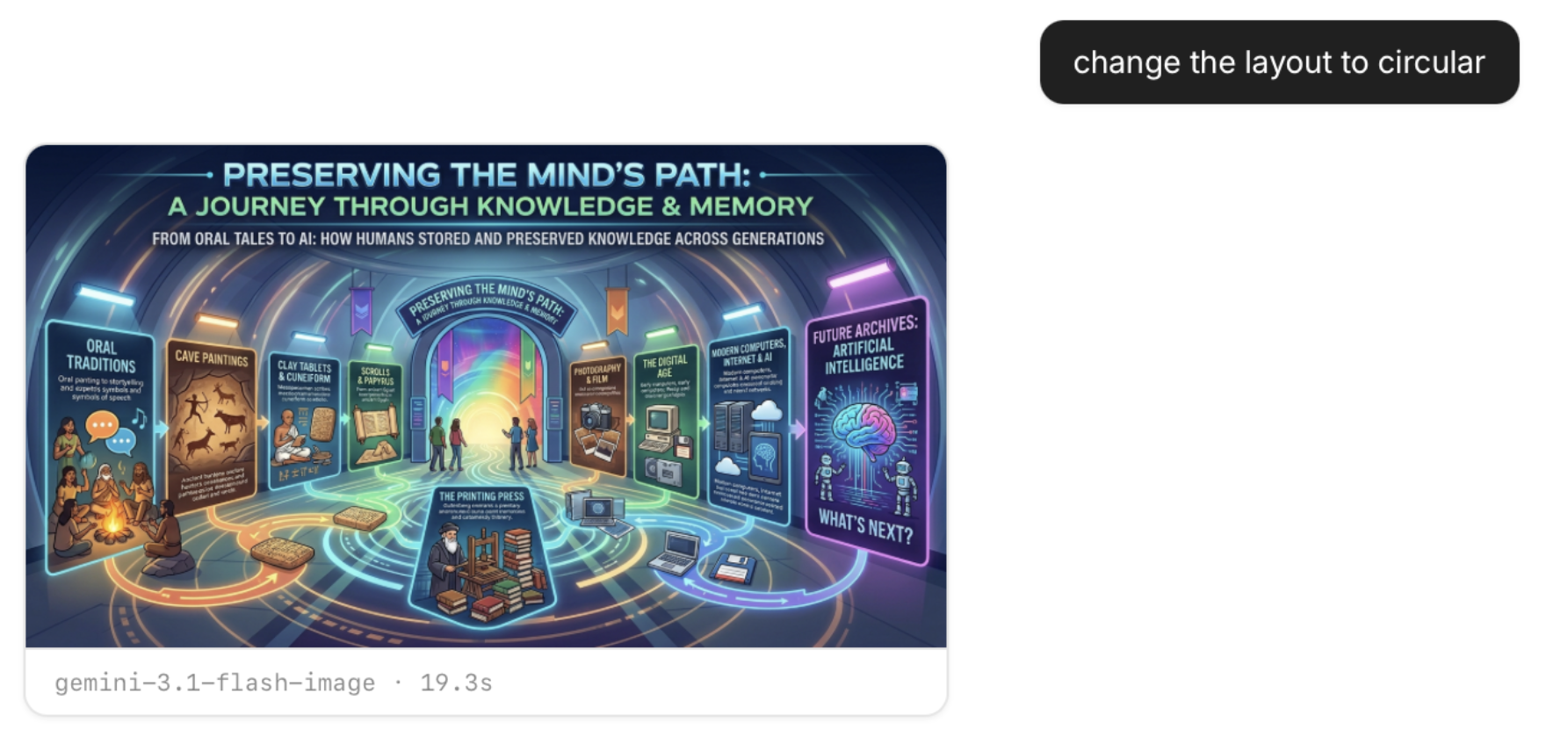}
        \subcaption{User edit request 1: ``change the layout to circular.'' The corridor survives and
        circular motifs are added as decoration instead: a global instruction
        lands as a local change, and the user has no way to say which parts of
        the image the phrase ``circular layout'' is meant to govern.}
        \label{fig:gemini-circular}
        \Description{The same corridor scene regenerated with curved arcs added
        to the floor and ceiling; the panel arrangement is essentially unchanged.}
    \end{subfigure}
    \\[1ex]
    \begin{subfigure}{\columnwidth}
        \includegraphics[width=0.6\columnwidth]{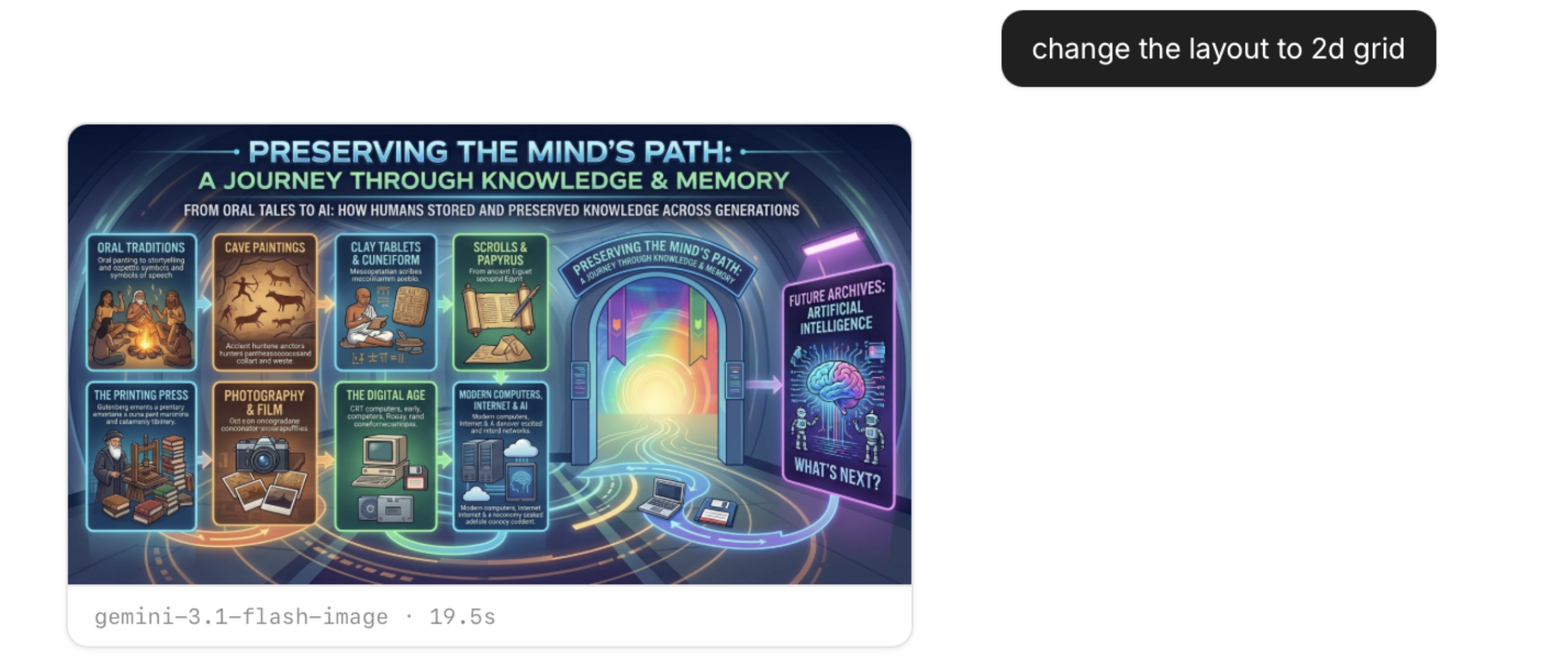}
        \subcaption{User edit request 2: ``change the layout to 2d grid.'' Here the arrangement does
        change, but the image is re-synthesized around it and the panel contents,
        labels, and color drift with it--the user cannot hold the content fixed
        while varying the layout.}
        \label{fig:gemini-grid}
        \Description{The same brief rendered as a two-by-four grid of exhibit
        panels; the illustrations and labels inside the panels differ from the
        previous two images.}
    \end{subfigure}
    \caption{Three turns of one Gemini chat session on the same infographic brief. Layout intent is expressible only as prose, so an edit either
    under-applies (b) or carries the rest of the image with it (c), while every
    turn stays anchored to the initial sample (a).}
    \Description{Three chat-based attempts to change only an infographic layout. The initial image uses an isometric corridor; a request for a circular layout mostly adds circular decoration without changing the structure; a request for a two-dimensional grid changes the layout but also changes content, labels, and color.}
    \label{fig:gemini-iterations}
\end{figure}

%% file: sections/22_appendix_c.tex
\section{System Prompts}\label{sec:appendix-prompts}

\subsection{Plan Generation}
\begin{lstlisting}[style=prompt]
You are an expert AI Image Planner. Turn the user's idea into a structured, editable Image Plan -- NOT a finished prompt, and NOT an image.

FIRST, silently infer the TASK TYPE (e.g. photo, illustration, character asset, icon set, logo, poster, infographic, UI asset, product mockup, diagram). Then plan in the TERMS THAT MATTER for that task. Do NOT force a photographic frame -- camera angle, lighting direction, depth of field, time of day -- onto tasks that don't have one (a flat infographic, icon set, logo, or UI asset has no "lighting" or "camera").

Output the plan in EXACTLY this structure (it is parsed downstream):

1. Concept Analysis
    * Intent: <one sentence -- the core purpose of the image>
    * Audience: <who it is for>
    * Mood: <2-5 words>

2. Structural Plan
   List ONLY the dimensions that are meaningful for the inferred task, each as a "**Label:** value" bullet. Choose from -- but do not force -- these dimensions:
    * Subject -- the main entity/entities and their key features.
    * Composition / Layout -- arrangement, framing, spatial structure (for graphic tasks: how panels/elements are laid out).
    * Background -- setting or backdrop.
    * Lighting -- ONLY for photographic / rendered 3D scenes; omit for flat/graphic tasks.
    * Color / Palette -- dominant and accent colors.
    * Style -- medium, treatment, aesthetic.
    * Typography / Text -- only if the image contains lettering.

Hard rules:
* Begin directly with the heading "1. Concept Analysis". NEVER include preamble, acknowledgments, or restatements (no "Certainly!", "Here's...", "Sure!").
* Keep each field to 1-2 sentences. Use bullets and bold labels for scannability.
* OMIT dimensions that don't apply to the task rather than inventing generic content for them.
* The project's output aspect ratio is \$\{outputAspect\}. If you mention an aspect ratio or framing in Composition / Layout, it MUST be \$\{outputAspect\}. Never introduce a different ratio.
\end{lstlisting}

\subsection{Decomposition into Plan Elements}
\label{sec:appendix-decompose-prompt}
\begin{lstlisting}[style=prompt]
You are the system's planning agent.

Your job is to turn a user's image request (given as a structured Image Plan in markdown) into a set of editable control points for a node-graph editor.

Do NOT use a fixed schema. Do NOT blindly include subject, background, lighting, color, or camera. Only include a control point if it is meaningful for THIS specific image task.

First infer the image task type. Then decide which control dimensions are useful for that task. Then propose user-facing control points.

A good control point is:
- meaningful for the current image task
- likely to affect generation
- independently editable by the user
- useful for future locking, variation, or refinement

A bad control point is:
- generic but irrelevant
- only included because a fixed schema expects it
- too vague to edit
- redundant with another control point

Also list common controls you intentionally omitted and why.

Output ONLY a single JSON object (no preamble, no markdown fences, no explanation). Schema:

{
  "inferredTaskType": one of "single\_image" | "character\_asset" | "asset\_set" | "icon\_set" | "sticker" | "poster" | "background\_scene" | "product\_mockup" | "infographic" | "ui\_asset" | "logo" | "reference\_variation\_set" | "other",
  "taskRationale": "one sentence on why this task type",
  "agentMessage": "one short friendly sentence: which controls you found and which you skipped",
  "controlPoints": [
    {
      "label": "short human label",
      "role": one of "identity" | "appearance" | "composition" | "pose\_action" | "environment" | "style" | "color" | "lighting" | "camera" | "material" | "text" | "constraint" | "negative\_constraint" | "variation\_axis" | "consistency\_rule" | "export\_format" | "other",
      "value": "1-3 sentences suitable for inclusion in an image-generation prompt",
      "description": "one sentence on what this control governs",
      "relevance": "essential" | "useful" | "optional",
      "granularity": "broad" | "specific"
    }
  ],
  "omittedControlPoints": [
    { "label": "Complex background scene", "reason": "This is a reusable asset, so a detailed scene would reduce reusability." }
  ],
  "suggestedAspectRatio": "1:1" | "4:3" | "3:2" | "16:9" | "9:16",
  "steps": [
    { "label": "short step title", "caption": "one short sentence explaining this step", "icon": "a single emoji or 1-2 word icon hint" }
  ]
}

Rules:
- Propose 3-8 control points based on SEMANTIC RELEVANCE, not a target count.
- Do not force irrelevant controls. Do list what you omitted.
- ONLY when inferredTaskType is "infographic": also return an ordered "steps" array (3-8 items) describing the sequence the infographic depicts, each with a short label and a one-sentence caption. For EVERY other task type, OMIT "steps" entirely (do not return an empty array).
- Output valid JSON only. Do not wrap in markdown code fences.
\end{lstlisting}

\subsection{Conversational Agent}
\label{sec:appendix-prompt-agent}
\begin{lstlisting}[style=prompt]
You are editing a semantic image-generation graph. You do not generate images. You do not write the final Gemini prompt. Your job is to propose structured edits to the graph so the existing Gemini compiler can later generate the final image.

You will be given the current graph (nodes with id, type, title, text, influence 0--100, constraint mode, priority rank) and a natural-language request from the user. Palette nodes additionally list their structured swatches (hex colors). Translate the request into edits that span as many nodes as needed to achieve the intent.

Output ONLY a single JSON object (no preamble, no markdown fences, no explanation) of this shape:

\{
  "summary": "one short sentence describing the overall change",
  "rationale": "1-3 sentences explaining why these edits achieve the request",
  "operations": [ /* see operation shapes below */ ]
\}
\end{lstlisting}

\begin{lstlisting}[style=prompt]
Operation shapes (include EXACTLY the listed keys; every operation MUST include a non-empty "reason"):
- \{ "op": "update\_node\_text", "nodeId": "...", "field": "text" | "title" | "compiledSummary", "after": "new value", "reason": "..." \}
- \{ "op": "set\_node\_influence", "nodeId": "...", "after": 0-100, "reason": "..." \}
- \{ "op": "set\_constraint\_mode", "nodeId": "...", "after": "off" | "loose" | "guide" | "strong" | "locked", "reason": "..." \}
- \{ "op": "set\_priority\_rank", "nodeId": "...", "after": 1, "reason": "..." \}
- \{ "op": "set\_palette\_colors", "nodeId": "...", "after": [ \{ "hex": "#rrggbb", "label": "optional name" \} ], "reason": "..." \}
- \{ "op": "add\_node", "node": \{ "id": "new-kebab-id", "type": "subject" | "composition" | "background" | "lighting" | "palette" | "style" | "typography" | "details", "title": "...", "text": "...", "influence": 0-100 \}, "reason": "..." \}
- \{ "op": "remove\_node", "nodeId": "...", "reason": "..." \}
- \{ "op": "connect\_nodes", "sourceId": "...", "targetId": "...", "edgeType": "contains" | "informs" | "overrides" | "refines", "reason": "..." \}
- \{ "op": "disconnect\_nodes", "sourceId": "...", "targetId": "...", "reason": "..." \}

Rules:
- Only edit nodes marked editable=true in the graph snapshot.
- To recolor a palette node, use set\_palette\_colors with concrete hex values (this updates the actual swatches and the compiled palette condition); do NOT try to recolor it via update\_node\_text.
- Never remove the last subject or composition node unless the user explicitly asks to.
- Prefer editing text / influence / mode / priority / palette colors over adding or removing nodes.
- Keep influence in 0--100 and constraint modes within the allowed set.
- Do not invent node ids that aren't in the graph (except for add\_node, which introduces a new id).
- Output valid JSON only.
\end{lstlisting}

\subsection{Final Image Generation}
\label{app:final_prompt}

\begin{lstlisting}[style=prompt]
=== SYSTEM INSTRUCTION ===
Before generating the image, first state your plan for it as plain text, using EXACTLY these labels -- each on its own line, in this order -- then generate the final image:
SUBJECT: <the main subject and how you will render it>
COMPOSITION: <layout, framing, and placement>
LIGHTING: <light direction, mood, atmosphere>
PALETTE: <dominant and accent colors>
WILL: <key things you will include>
WONT: <things you will deliberately avoid>
Keep each line to one sentence. After the plan, output the single final image.

You are a node compiler: a layer that synthesizes semantic nodes into a single final image. Create one final image by composing the semantic nodes provided in the node manifest.

Overall concept / goal: Create a 1950s retro sci-fi magazine cover featuring a Martian city and astronauts.

OUTPUT FORMAT:
* Aspect ratio: 1:1
* Image style target: 1950s pulp sci-fi magazine cover illustration style, mimicking hand-painted gouache or tempera on paper. Features visible halftone dots, subtle paper texture, and slight ink-bleed artifacts.
* Final image should feel coherent, intentional, and polished.
* Do not make a collage or moodboard. Use the references to synthesize one unified image.

Each manifest node declares an id, type, priority, an influence/constraint level, the semantic role its attached reference image should (and should not) be used for, and a text specification. Use each reference only for its declared semantic role, and honor each node in proportion to its stated constraint level.

PRIORITY CONSTRAINTS (obey these above all, strongest first):
- [STRONG] main subject (subject): Astronauts in the foreground wearing bulky, studded 1950s-style spacesuits with reflective helmet visors, looking out towards the city. -- honor closely.
- [STRONG] composition/layout (composition): Classic pulp magazine cover layout. A bold title spans the top, the main illustration fills the center, and the astronauts are positioned in the lower third to establish scale against the towering city. -- honor closely.
- [STRONG] scene/background (scene): A futuristic Martian city with tall pink crystal towers and silver plate-like flying saucers docking on elevated platforms, set against a dramatic deep cobalt blue sky and orange-brown terrain. -- honor closely.
- [STRONG] color palette (palette): #0047ab, #c05a2b, #ff69b4, #c0c0c0, cobalt (#1f4ea8), blue (#3a6cb0), orange (#e08a4a), brown (#7a5a3a), pink (#d97aae) -- honor closely.

HARD RULES (these are absolute -- they override every node above):
- Aspect Ratio: Must be exactly 1:1.
- Style: Avoid modern 3D digital rendering; must look like a physical, mid-century printed magazine cover.
- Text Style: The main title must have slightly ink-bleed edges and use bold serif capital letters.

SYNTHESIS INSTRUCTION: Generate a single final image that follows the node manifest. Use each reference only for its declared semantic role. Resolve conflicts according to priority and constraint level -- locked constraints win over strong, strong over guide. The final result should look like a naturally unified image, not a pasted combination of parts.

=== INTERLEAVED CONTENTS ===
NODE MANIFEST:
[NODE 1] id: subject_01 type: main subject priority: high influence: 78/100 (strong)
use attached image for: subject identity, silhouette, proportions, key visual features, pose, clothing/material details.
do not use attached image for: background, lighting, camera framing, or color grading unless explicitly repeated below.
text specification: Astronauts in the foreground wearing bulky, studded 1950s-style spacesuits with reflective helmet visors, looking out towards the city.
This is an important constraint -- honor it closely.
[IMAGE: subject_01]
  <base64-encoded reference image (PNG), elided>
[NODE 2] id: composition_01 type: composition/layout priority: high influence: 78/100 (strong)
use attached image for: spatial layout, framing, camera angle, relative object placement, negative space, and major visual balance.
do not use attached image for: subject identity, color palette, rendering style, lighting, or background details unless explicitly repeated below.
text specification: Classic pulp magazine cover layout. A bold title spans the top, the main illustration fills the center, and the astronauts are positioned in the lower third to establish scale against the towering city.
This is an important constraint -- honor it closely.
[IMAGE: composition_01]
  <base64-encoded reference image (PNG), elided>
[NODE 3] id: scene_01 type: scene/background priority: medium influence: 78/100 (strong)
use attached image for: environment type, architectural/natural setting, background mood, spatial depth.
do not use attached image for: subject identity or exact composition unless explicitly stated.
text specification: A futuristic Martian city with tall pink crystal towers and silver plate-like flying saucers docking on elevated platforms, set against a dramatic deep cobalt blue sky and orange-brown terrain.
This is an important constraint -- honor it closely.
[IMAGE: scene_01]
  <base64-encoded reference image (PNG), elided>
[NODE 4] id: palette_01 type: color palette priority: medium influence: 78/100 (strong)
text specification: #0047ab, #c05a2b, #ff69b4, #c0c0c0, cobalt (#1f4ea8), blue (#3a6cb0), orange (#e08a4a), brown (#7a5a3a), pink (#d97aae)
This is an important constraint -- honor it closely.
(no attached image -- apply the text specification only)
[NODE 5] id: lighting_01 type: lighting priority: medium influence: 50/100 (guide)
use attached image for: light direction, shadow softness, contrast, atmosphere, time-of-day feeling.
do not use attached image for: objects, scene, or subject identity.
text specification: High-contrast retro illustration lighting. A strong, dramatic key light casts long shadows across the terrain, with bright, specular reflections on the silver saucers and helmet visors.
[IMAGE: lighting_01]
  <base64-encoded reference image (PNG), elided>
\end{lstlisting}

%% file: main.bbl

\begin{thebibliography}{36}


\ifx \showCODEN    \undefined \def \showCODEN     #1{\unskip}     \fi
\ifx \showISBNx    \undefined \def \showISBNx     #1{\unskip}     \fi
\ifx \showISBNxiii \undefined \def \showISBNxiii  #1{\unskip}     \fi
\ifx \showISSN     \undefined \def \showISSN      #1{\unskip}     \fi
\ifx \showLCCN     \undefined \def \showLCCN      #1{\unskip}     \fi
\ifx \shownote     \undefined \def \shownote      #1{#1}          \fi
\ifx \showarticletitle \undefined \def \showarticletitle #1{#1}   \fi
\ifx \showURL      \undefined \def \showURL       {\relax}        \fi
\providecommand\bibfield[2]{#2}
\providecommand\bibinfo[2]{#2}
\providecommand\natexlab[1]{#1}
\providecommand\showeprint[2][]{arXiv:#2}

\bibitem[Angert et~al\mbox{.}(2023)]%
        {angert2023spellburst}
\bibfield{author}{\bibinfo{person}{Tyler Angert}, \bibinfo{person}{Miroslav
  Suzara}, \bibinfo{person}{Jenny Han}, \bibinfo{person}{Christopher Pondoc},
  {and} \bibinfo{person}{Hariharan Subramonyam}.}
  \bibinfo{year}{2023}\natexlab{}.
\newblock \showarticletitle{Spellburst: A Node-based Interface for Exploratory
  Creative Coding with Natural Language Prompts}.
\newblock \bibinfo{journal}{\emph{UIST}} (\bibinfo{date}{Oct.}
  \bibinfo{year}{2023}), \bibinfo{pages}{1–22}.
\newblock
\href{https://doi.org/10.1145/3586183.3606719}{doi:\nolinkurl{10.1145/3586183.3606719}}


\bibitem[Arifin and Hasbi(2025)]%
        {arifin2025illustration}
\bibfield{author}{\bibinfo{person}{Irfan Arifin} {and} \bibinfo{person}{Hasbi
  Hasbi}.} \bibinfo{year}{2025}\natexlab{}.
\newblock \showarticletitle{Illustration process in visual communication: a
  qualitative study on storybook product development}.
\newblock \bibinfo{journal}{\emph{International Journal of Visual and
  Performing Arts}} \bibinfo{volume}{7}, \bibinfo{number}{1}
  (\bibinfo{year}{2025}), \bibinfo{pages}{195--206}.
\newblock
\showISSN{2684-9259}
\href{https://doi.org/10.31763/viperarts.v7i1.1673}{doi:\nolinkurl{10.31763/viperarts.v7i1.1673}}


\bibitem[ComfyAnonymous and Contributors(2023)]%
        {ComfyUI}
\bibfield{author}{\bibinfo{person}{ComfyAnonymous} {and}
  \bibinfo{person}{Contributors}.} \bibinfo{year}{2023}\natexlab{}.
\newblock \bibinfo{booktitle}{\emph{ComfyUI: A modular stable diffusion GUI and
  backend}}.
\newblock
\urldef\tempurl%
\url{https://github.com/comfy-org/comfyui}
\showURL{%
\tempurl}


\bibitem[{Design Council}(2005)]%
        {designcouncil2005}
\bibfield{author}{\bibinfo{person}{{Design Council}}.}
  \bibinfo{year}{2005}\natexlab{}.
\newblock \bibinfo{title}{A study of the design process -- The Double Diamond}.
\newblock
\urldef\tempurl%
\url{http://www.designcouncil.org.uk/sites/default/files/asset/document/ElevenLessons_Design_Council%20%282%29.pdf}
\showURL{%
\tempurl}


\bibitem[Dorst and Cross(2001)]%
        {Dorst2001Creativity}
\bibfield{author}{\bibinfo{person}{Kees Dorst} {and} \bibinfo{person}{Nigel
  Cross}.} \bibinfo{year}{2001}\natexlab{}.
\newblock \showarticletitle{Creativity in the design process: co-evolution of
  problem–solution}.
\newblock \bibinfo{journal}{\emph{Design Studies}} \bibinfo{volume}{22},
  \bibinfo{number}{5} (\bibinfo{year}{2001}), \bibinfo{pages}{425--437}.
\newblock
\href{https://doi.org/10.1016/S0142-694X(01)00009-6}{doi:\nolinkurl{10.1016/S0142-694X(01)00009-6}}


\bibitem[{Google DeepMind}(2025)]%
        {Gemini3ProImage}
\bibfield{author}{\bibinfo{person}{{Google DeepMind}}.}
  \bibinfo{year}{2025}\natexlab{}.
\newblock \bibinfo{booktitle}{\emph{Gemini 3 Pro Image}}.
\newblock
\shownote{Image model}.
\newblock
\urldef\tempurl%
\url{https://ai.google.dev/gemini-api/docs/models/gemini-3-pro-image}
\showURL{%
\tempurl}


\bibitem[{Google DeepMind}(2026a)]%
        {Gemini31FlashImage}
\bibfield{author}{\bibinfo{person}{{Google DeepMind}}.}
  \bibinfo{year}{2026}\natexlab{a}.
\newblock \bibinfo{booktitle}{\emph{Gemini 3.1 Flash Image}}.
\newblock
\shownote{Image model}.
\newblock
\urldef\tempurl%
\url{https://ai.google.dev/gemini-api/docs/models/gemini-3.1-flash-image}
\showURL{%
\tempurl}


\bibitem[{Google DeepMind}(2026b)]%
        {Gemini35Flash}
\bibfield{author}{\bibinfo{person}{{Google DeepMind}}.}
  \bibinfo{year}{2026}\natexlab{b}.
\newblock \bibinfo{booktitle}{\emph{Gemini 3.5 Flash}}.
\newblock
\shownote{Large language model}.
\newblock
\urldef\tempurl%
\url{https://deepmind.google/models/model-cards/gemini-3-5-flash/}
\showURL{%
\tempurl}


\bibitem[Hao et~al\mbox{.}(2023)]%
        {hao2023optimizingpromptstexttoimagegeneration}
\bibfield{author}{\bibinfo{person}{Yaru Hao}, \bibinfo{person}{Zewen Chi},
  \bibinfo{person}{Li Dong}, {and} \bibinfo{person}{Furu Wei}.}
  \bibinfo{year}{2023}\natexlab{}.
\newblock \showarticletitle{Optimizing Prompts for Text-to-Image Generation}.
  In \bibinfo{booktitle}{\emph{Advances in Neural Information Processing
  Systems}}, \bibfield{editor}{\bibinfo{person}{A.~Oh},
  \bibinfo{person}{T.~Naumann}, \bibinfo{person}{A.~Globerson},
  \bibinfo{person}{K.~Saenko}, \bibinfo{person}{M.~Hardt}, {and}
  \bibinfo{person}{S.~Levine}} (Eds.), Vol.~\bibinfo{volume}{36}.
  \bibinfo{publisher}{Curran Associates, Inc.}, \bibinfo{pages}{66923--66939}.
\newblock
\href{https://doi.org/10.52202/075280-2923}{doi:\nolinkurl{10.52202/075280-2923}}


\bibitem[He et~al\mbox{.}(2025a)]%
        {he2025plangen}
\bibfield{author}{\bibinfo{person}{Runze He}, \bibinfo{person}{Bo cheng},
  \bibinfo{person}{Yuhang Ma}, \bibinfo{person}{Qingxiang Jia},
  \bibinfo{person}{Shanyuan Liu}, \bibinfo{person}{Ao Ma},
  \bibinfo{person}{Xiaoyu Wu}, \bibinfo{person}{Liebucha Wu},
  \bibinfo{person}{Dawei Leng}, {and} \bibinfo{person}{Yuhui Yin}.}
  \bibinfo{year}{2025}\natexlab{a}.
\newblock \showarticletitle{PlanGen: Towards Unified Layout Planning and Image
  Generation in Auto-Regressive Vision Language Models}. In
  \bibinfo{booktitle}{\emph{2025 IEEE/CVF International Conference on Computer
  Vision (ICCV)}}. \bibinfo{pages}{18143--18154}.
\newblock
\href{https://doi.org/10.1109/ICCV51701.2025.01686}{doi:\nolinkurl{10.1109/ICCV51701.2025.01686}}


\bibitem[He et~al\mbox{.}(2025b)]%
        {he2025automatedblackboxpromptengineering}
\bibfield{author}{\bibinfo{person}{Yutong He}, \bibinfo{person}{Alexander
  Robey}, \bibinfo{person}{Naoki Murata}, \bibinfo{person}{Yiding Jiang},
  \bibinfo{person}{Joshua~Nathaniel Williams}, \bibinfo{person}{George~J.
  Pappas}, \bibinfo{person}{Hamed Hassani}, \bibinfo{person}{Yuki Mitsufuji},
  \bibinfo{person}{Ruslan Salakhutdinov}, {and} \bibinfo{person}{J~Zico
  Kolter}.} \bibinfo{year}{2025}\natexlab{b}.
\newblock \showarticletitle{Automated Black-box Prompt Engineering for
  Personalized Text-to-Image Generation}.
\newblock \bibinfo{journal}{\emph{Transactions on Machine Learning Research}}
  (\bibinfo{year}{2025}).
\newblock
\showISSN{2835-8856}
\urldef\tempurl%
\url{https://openreview.net/forum?id=IVYVDN6pJ6}
\showURL{%
\tempurl}


\bibitem[Hegemann et~al\mbox{.}(2026)]%
        {hegemann2026tomigointerpretabledesignconcept}
\bibfield{author}{\bibinfo{person}{Lena Hegemann}, \bibinfo{person}{Xinyi Wen},
  \bibinfo{person}{Michael~A. Hedderich}, \bibinfo{person}{Tarmo Nurmi}, {and}
  \bibinfo{person}{Hariharan Subramonyam}.} \bibinfo{year}{2026}\natexlab{}.
\newblock \showarticletitle{ToMigo: Interpretable Design Concept Graphs for
  Aligning Generative AI with Creative Intent}. In
  \bibinfo{booktitle}{\emph{Proceedings of the 2026 Designing Interactive
  Systems Conference}} \emph{(\bibinfo{series}{DIS '26})}.
  \bibinfo{publisher}{Association for Computing Machinery},
  \bibinfo{address}{New York, NY, USA}, \bibinfo{pages}{643–660}.
\newblock
\showISBNx{9798400725630}
\href{https://doi.org/10.1145/3800645.3813064}{doi:\nolinkurl{10.1145/3800645.3813064}}


\bibitem[Huang et~al\mbox{.}(2025)]%
        {sketchgpt}
\bibfield{author}{\bibinfo{person}{Zeyuan Huang}, \bibinfo{person}{Cangjun
  Gao}, \bibinfo{person}{Yaxian Shan}, \bibinfo{person}{Haoxiang Hu},
  \bibinfo{person}{Qingkun Li}, \bibinfo{person}{Xiaoming Deng},
  \bibinfo{person}{Cuixia Ma}, \bibinfo{person}{Yu-Kun Lai},
  \bibinfo{person}{Yong-Jin Liu}, \bibinfo{person}{Feng Tian},
  \bibinfo{person}{Guozhong Dai}, {and} \bibinfo{person}{Hongan Wang}.}
  \bibinfo{year}{2025}\natexlab{}.
\newblock \showarticletitle{SketchGPT: A Sketch-based Multimodal Interface for
  Application-Agnostic LLM Interaction}. In
  \bibinfo{booktitle}{\emph{Proceedings of the 38th Annual ACM Symposium on
  User Interface Software and Technology}} \emph{(\bibinfo{series}{UIST '25})}.
  \bibinfo{publisher}{Association for Computing Machinery},
  \bibinfo{address}{New York, NY, USA}, Article \bibinfo{articleno}{157},
  \bibinfo{numpages}{18}~pages.
\newblock
\showISBNx{9798400720376}
\href{https://doi.org/10.1145/3746059.3747598}{doi:\nolinkurl{10.1145/3746059.3747598}}


\bibitem[Ji et~al\mbox{.}(2025)]%
        {ji2026instructionbasedimageeditingplanning}
\bibfield{author}{\bibinfo{person}{Liya Ji}, \bibinfo{person}{Chenyang Qi},
  {and} \bibinfo{person}{Qifeng Chen}.} \bibinfo{year}{2025}\natexlab{}.
\newblock \showarticletitle{Instruction-based Image Editing with Planning,
  Reasoning, and Generation}. In \bibinfo{booktitle}{\emph{Proceedings of the
  IEEE/CVF International Conference on Computer Vision (ICCV)}}.
  \bibinfo{pages}{17506--17515}.
\newblock


\bibitem[Jiang et~al\mbox{.}(2024)]%
        {jiang2025selfplanningcodegenerationlarge}
\bibfield{author}{\bibinfo{person}{Xue Jiang}, \bibinfo{person}{Yihong Dong},
  \bibinfo{person}{Lecheng Wang}, \bibinfo{person}{Zheng Fang},
  \bibinfo{person}{Qiwei Shang}, \bibinfo{person}{Ge Li}, \bibinfo{person}{Zhi
  Jin}, {and} \bibinfo{person}{Wenpin Jiao}.} \bibinfo{year}{2024}\natexlab{}.
\newblock \showarticletitle{Self-Planning Code Generation with Large Language
  Models}.
\newblock \bibinfo{journal}{\emph{ACM Trans. Softw. Eng. Methodol.}}
  \bibinfo{volume}{33}, \bibinfo{number}{7}, Article \bibinfo{articleno}{182}
  (\bibinfo{date}{Sept.} \bibinfo{year}{2024}), \bibinfo{numpages}{30}~pages.
\newblock
\showISSN{1049-331X}
\href{https://doi.org/10.1145/3672456}{doi:\nolinkurl{10.1145/3672456}}


\bibitem[Kuba and Jeong(2024)]%
        {Kuba02042024}
\bibfield{author}{\bibinfo{person}{Renata Kuba} {and} \bibinfo{person}{Allan
  Jeong}.} \bibinfo{year}{2024}\natexlab{}.
\newblock \showarticletitle{Sequential analysis of design processes performed
  by design experts and college students in infographic design}.
\newblock \bibinfo{journal}{\emph{Journal of Visual Literacy}}
  \bibinfo{volume}{43}, \bibinfo{number}{2} (\bibinfo{year}{2024}),
  \bibinfo{pages}{124--146}.
\newblock
\href{https://doi.org/10.1080/1051144X.2024.2349380}{doi:\nolinkurl{10.1080/1051144X.2024.2349380}}


\bibitem[Kyaw and Sivalingam(2025)]%
        {kyaw2025nodebasededitingmultimodalgeneration}
\bibfield{author}{\bibinfo{person}{Alexander~Htet Kyaw} {and}
  \bibinfo{person}{Lenin~Ravindranath Sivalingam}.}
  \bibinfo{year}{2025}\natexlab{}.
\newblock \bibinfo{title}{Node-Based Editing for Multimodal Generation of Text,
  Audio, Image, and Video}.
\newblock
\showeprint[arxiv]{2511.03227}~[cs.HC]
\urldef\tempurl%
\url{https://arxiv.org/abs/2511.03227}
\showURL{%
\tempurl}


\bibitem[Li et~al\mbox{.}(2024)]%
        {Li_2024_CVPR_genaibench}
\bibfield{author}{\bibinfo{person}{Baiqi Li}, \bibinfo{person}{Zhiqiu Lin},
  \bibinfo{person}{Deepak Pathak}, \bibinfo{person}{Jiayao Li},
  \bibinfo{person}{Yixin Fei}, \bibinfo{person}{Kewen Wu},
  \bibinfo{person}{Xide Xia}, \bibinfo{person}{Pengchuan Zhang},
  \bibinfo{person}{Graham Neubig}, {and} \bibinfo{person}{Deva Ramanan}.}
  \bibinfo{year}{2024}\natexlab{}.
\newblock \showarticletitle{Evaluating and Improving Compositional
  Text-to-Visual Generation}. In \bibinfo{booktitle}{\emph{Proceedings of the
  IEEE/CVF Conference on Computer Vision and Pattern Recognition (CVPR)
  Workshops}}. \bibinfo{pages}{5290--5301}.
\newblock


\bibitem[Li et~al\mbox{.}(2026)]%
        {li2026qwen_image_bench}
\bibfield{author}{\bibinfo{person}{Niantong Li}, \bibinfo{person}{Guangzheng
  Hu}, \bibinfo{person}{Weixu Qiao}, \bibinfo{person}{Ying Ba},
  \bibinfo{person}{Qichen Hong}, \bibinfo{person}{Shijun Shen},
  \bibinfo{person}{Jinlin Wang}, \bibinfo{person}{Fan Zhou},
  \bibinfo{person}{Jianye Kang}, \bibinfo{person}{Xin Shang},
  \bibinfo{person}{Ziyi He}, \bibinfo{person}{Wei Wang}, \bibinfo{person}{Dalin
  Li}, \bibinfo{person}{Jiahao Li}, \bibinfo{person}{Jie Zhang},
  \bibinfo{person}{Kaiyuan Gao}, \bibinfo{person}{Kun Yan},
  \bibinfo{person}{Lihan Jiang}, \bibinfo{person}{Ningyuan Tang},
  \bibinfo{person}{Shengming Yin}, \bibinfo{person}{Tianhe Wu},
  \bibinfo{person}{Xiao Xu}, \bibinfo{person}{Xiaoyue Chen},
  \bibinfo{person}{Yuxiang Chen}, \bibinfo{person}{Yan Shu},
  \bibinfo{person}{Yanran Zhang}, \bibinfo{person}{Yilei Chen},
  \bibinfo{person}{Yixian Xu}, \bibinfo{person}{Zekai Zhang},
  \bibinfo{person}{Zhendong Wang}, \bibinfo{person}{Zihao Liu},
  \bibinfo{person}{Zikai Zhou}, \bibinfo{person}{Hongzhu Shi},
  \bibinfo{person}{Yi Wang}, \bibinfo{person}{Bing Zhao}, \bibinfo{person}{Hu
  Wei}, \bibinfo{person}{Lin Qu}, {and} \bibinfo{person}{Chenfei Wu}.}
  \bibinfo{year}{2026}\natexlab{}.
\newblock \showarticletitle{Qwen-Image-Bench: From Generation to Creation in
  Text-to-Image Evaluation}.
\newblock \bibinfo{journal}{\emph{arXiv preprint arXiv: 2605.28091}}
  (\bibinfo{year}{2026}).
\newblock


\bibitem[Liu and Chilton(2022)]%
        {10.1145/3491102.3501825}
\bibfield{author}{\bibinfo{person}{Vivian Liu} {and} \bibinfo{person}{Lydia~B
  Chilton}.} \bibinfo{year}{2022}\natexlab{}.
\newblock \showarticletitle{Design Guidelines for Prompt Engineering
  Text-to-Image Generative Models}. In \bibinfo{booktitle}{\emph{Proceedings of
  the 2022 CHI Conference on Human Factors in Computing Systems}} (New Orleans,
  LA, USA) \emph{(\bibinfo{series}{CHI '22})}. \bibinfo{publisher}{Association
  for Computing Machinery}, \bibinfo{address}{New York, NY, USA}, Article
  \bibinfo{articleno}{384}, \bibinfo{numpages}{23}~pages.
\newblock
\showISBNx{9781450391573}
\href{https://doi.org/10.1145/3491102.3501825}{doi:\nolinkurl{10.1145/3491102.3501825}}


\bibitem[Ma et~al\mbox{.}(2025)]%
        {ma2025artkit}
\bibfield{author}{\bibinfo{person}{Jiaju Ma}, \bibinfo{person}{Chau Vu},
  \bibinfo{person}{Asya Lyubavina}, \bibinfo{person}{Catherine Liu}, {and}
  \bibinfo{person}{Jingyi Li}.} \bibinfo{year}{2025}\natexlab{}.
\newblock \showarticletitle{Computational Scaffolding of Composition, Value,
  and Color for Disciplined Drawing}. In \bibinfo{booktitle}{\emph{Proceedings
  of the 38th Annual ACM Symposium on User Interface Software and Technology}}
  \emph{(\bibinfo{series}{UIST '25})}. \bibinfo{publisher}{Association for
  Computing Machinery}, \bibinfo{address}{New York, NY, USA}, Article
  \bibinfo{articleno}{161}, \bibinfo{numpages}{15}~pages.
\newblock
\showISBNx{9798400720376}
\href{https://doi.org/10.1145/3746059.3747605}{doi:\nolinkurl{10.1145/3746059.3747605}}


\bibitem[{MidJourney}(2026)]%
        {midjourney}
\bibfield{author}{\bibinfo{person}{{MidJourney}}.}
  \bibinfo{year}{2026}\natexlab{}.
\newblock \bibinfo{title}{{MidJourney}}.
\newblock
\urldef\tempurl%
\url{https://docs.midjourney.com}
\showURL{%
\tempurl}


\bibitem[Mou et~al\mbox{.}(2024)]%
        {mou2023dragondiffusionenablingdragstylemanipulation}
\bibfield{author}{\bibinfo{person}{Chong Mou}, \bibinfo{person}{Xintao Wang},
  \bibinfo{person}{Jiechong Song}, \bibinfo{person}{Ying Shan}, {and}
  \bibinfo{person}{Jian Zhang}.} \bibinfo{year}{2024}\natexlab{}.
\newblock \showarticletitle{DragonDiffusion: Enabling Drag-style Manipulation
  on Diffusion Models}. In \bibinfo{booktitle}{\emph{The Twelfth International
  Conference on Learning Representations}}.
\newblock
\urldef\tempurl%
\url{https://openreview.net/forum?id=OEL4FJMg1b}
\showURL{%
\tempurl}


\bibitem[Nguyen et~al\mbox{.}(2025)]%
        {nguyen2025genflowinteractivemodularimage}
\bibfield{author}{\bibinfo{person}{Duc-Hung Nguyen}, \bibinfo{person}{Huu-Phuc
  Huynh}, \bibinfo{person}{Minh-Triet Tran}, {and} \bibinfo{person}{Trung-Nghia
  Le}.} \bibinfo{year}{2025}\natexlab{}.
\newblock \bibinfo{title}{GenFlow: Interactive Modular System for Image
  Generation}.
\newblock
\showeprint[arxiv]{2506.21369}~[cs.CV]
\urldef\tempurl%
\url{https://arxiv.org/abs/2506.21369}
\showURL{%
\tempurl}


\bibitem[{OpenAI}(2026)]%
        {openai2026chatgpt}
\bibfield{author}{\bibinfo{person}{{OpenAI}}.} \bibinfo{year}{2026}\natexlab{}.
\newblock \bibinfo{title}{{ChatGPT}}.
\newblock
\urldef\tempurl%
\url{https://chatgpt.com}
\showURL{%
\tempurl}


\bibitem[Pang et~al\mbox{.}(2026)]%
        {pang2025interactivereasoningvisualizingcontrolling}
\bibfield{author}{\bibinfo{person}{Rock~Yuren Pang},
  \bibinfo{person}{K.~J.~Kevin Feng}, \bibinfo{person}{Shangbin Feng},
  \bibinfo{person}{Chu Li}, \bibinfo{person}{Weijia Shi},
  \bibinfo{person}{Yulia Tsvetkov}, \bibinfo{person}{Jeffrey Heer}, {and}
  \bibinfo{person}{Katharina Reinecke}.} \bibinfo{year}{2026}\natexlab{}.
\newblock \showarticletitle{Interactive Reasoning: Visualizing and Controlling
  Chain-of-Thought Reasoning in Large Language Models}. In
  \bibinfo{booktitle}{\emph{Proceedings of the 31st International Conference on
  Intelligent User Interfaces}} \emph{(\bibinfo{series}{IUI '26})}.
  \bibinfo{publisher}{Association for Computing Machinery},
  \bibinfo{address}{New York, NY, USA}, \bibinfo{pages}{852–867}.
\newblock
\showISBNx{9798400719844}
\href{https://doi.org/10.1145/3742413.3789091}{doi:\nolinkurl{10.1145/3742413.3789091}}


\bibitem[Sarukkai et~al\mbox{.}(2024)]%
        {sarukkai2024blockdetailscaffoldingsketchtoimage}
\bibfield{author}{\bibinfo{person}{Vishnu Sarukkai}, \bibinfo{person}{Lu Yuan},
  \bibinfo{person}{Mia Tang}, \bibinfo{person}{Maneesh Agrawala}, {and}
  \bibinfo{person}{Kayvon Fatahalian}.} \bibinfo{year}{2024}\natexlab{}.
\newblock \showarticletitle{Block and Detail: Scaffolding Sketch-to-Image
  Generation}. In \bibinfo{booktitle}{\emph{Proceedings of the 37th Annual ACM
  Symposium on User Interface Software and Technology}} (Pittsburgh, PA, USA)
  \emph{(\bibinfo{series}{UIST '24})}. \bibinfo{publisher}{Association for
  Computing Machinery}, \bibinfo{address}{New York, NY, USA}, Article
  \bibinfo{articleno}{33}, \bibinfo{numpages}{13}~pages.
\newblock
\showISBNx{9798400706288}
\href{https://doi.org/10.1145/3654777.3676444}{doi:\nolinkurl{10.1145/3654777.3676444}}


\bibitem[Sawyer(2021)]%
        {Sawyer2021Iterative}
\bibfield{author}{\bibinfo{person}{R.~Keith Sawyer}.}
  \bibinfo{year}{2021}\natexlab{}.
\newblock \showarticletitle{The iterative and improvisational nature of the
  creative process}.
\newblock \bibinfo{journal}{\emph{Journal of Creativity}} \bibinfo{volume}{31},
  \bibinfo{number}{3} (\bibinfo{year}{2021}), \bibinfo{pages}{100002}.
\newblock
\showISSN{2713-3745}
\href{https://doi.org/10.1016/j.yjoc.2021.100002}{doi:\nolinkurl{10.1016/j.yjoc.2021.100002}}


\bibitem[Shi et~al\mbox{.}(2025a)]%
        {10.1145/3706599.3720182}
\bibfield{author}{\bibinfo{person}{Xinyu Shi}, \bibinfo{person}{Shunan Guo},
  \bibinfo{person}{Jane Hoffswell}, \bibinfo{person}{Gromit Yeuk-Yin Chan},
  \bibinfo{person}{Victor~S. Bursztyn}, \bibinfo{person}{Jian Zhao}, {and}
  \bibinfo{person}{Eunyee Koh}.} \bibinfo{year}{2025}\natexlab{a}.
\newblock \showarticletitle{Comprehensive Sketching: Exploring Infographic
  Design Alternatives in Parallel}. In \bibinfo{booktitle}{\emph{Proceedings of
  the Extended Abstracts of the CHI Conference on Human Factors in Computing
  Systems}} \emph{(\bibinfo{series}{CHI EA '25})}.
  \bibinfo{publisher}{Association for Computing Machinery},
  \bibinfo{address}{New York, NY, USA}, Article \bibinfo{articleno}{145},
  \bibinfo{numpages}{8}~pages.
\newblock
\showISBNx{9798400713958}
\href{https://doi.org/10.1145/3706599.3720182}{doi:\nolinkurl{10.1145/3706599.3720182}}


\bibitem[Shi et~al\mbox{.}(2025b)]%
        {shi2025brickifyenablingexpressivedesign}
\bibfield{author}{\bibinfo{person}{Xinyu Shi}, \bibinfo{person}{Yinghou Wang},
  \bibinfo{person}{Ryan Rossi}, {and} \bibinfo{person}{Jian Zhao}.}
  \bibinfo{year}{2025}\natexlab{b}.
\newblock \showarticletitle{Brickify: Enabling Expressive Design Intent
  Specification through Direct Manipulation on Design Tokens}. In
  \bibinfo{booktitle}{\emph{Proceedings of the 2025 CHI Conference on Human
  Factors in Computing Systems}} \emph{(\bibinfo{series}{CHI '25})}.
  \bibinfo{publisher}{Association for Computing Machinery},
  \bibinfo{address}{New York, NY, USA}, Article \bibinfo{articleno}{424},
  \bibinfo{numpages}{20}~pages.
\newblock
\showISBNx{9798400713941}
\href{https://doi.org/10.1145/3706598.3714087}{doi:\nolinkurl{10.1145/3706598.3714087}}


\bibitem[Simone et~al\mbox{.}(2026)]%
        {desimone2026creooneshotimagegeneration}
\bibfield{author}{\bibinfo{person}{Zoe~De Simone}, \bibinfo{person}{Angie
  Boggust}, \bibinfo{person}{Fredo Durand}, \bibinfo{person}{Ashia Wilson},
  {and} \bibinfo{person}{Arvind Satyanarayan}.}
  \bibinfo{year}{2026}\natexlab{}.
\newblock \bibinfo{title}{Creo: From One-Shot Image Generation to Progressive,
  Co-Creative Ideation}.
\newblock
\showeprint[arxiv]{2604.13956}~[cs.HC]
\urldef\tempurl%
\url{https://arxiv.org/abs/2604.13956}
\showURL{%
\tempurl}


\bibitem[University(2004)]%
        {stanford}
\bibfield{author}{\bibinfo{person}{Stanford University}.}
  \bibinfo{year}{2004}\natexlab{}.
\newblock \bibinfo{title}{{M}{E} 113 --- web.stanford.edu}.
\newblock
  \bibinfo{howpublished}{\url{https://web.stanford.edu/class/me113/d\_thinking.html}}.
\newblock
\shownote{[Accessed 10-09-2026]}.
\newblock


\bibitem[Wadinambiarachchi et~al\mbox{.}(2024)]%
        {Wadinambiarachchi_2024}
\bibfield{author}{\bibinfo{person}{Samangi Wadinambiarachchi},
  \bibinfo{person}{Ryan~M. Kelly}, \bibinfo{person}{Saumya Pareek},
  \bibinfo{person}{Qiushi Zhou}, {and} \bibinfo{person}{Eduardo Velloso}.}
  \bibinfo{year}{2024}\natexlab{}.
\newblock \showarticletitle{The Effects of Generative AI on Design Fixation and
  Divergent Thinking}. In \bibinfo{booktitle}{\emph{Proceedings of the CHI
  Conference on Human Factors in Computing Systems}}
  \emph{(\bibinfo{series}{CHI ’24})}. \bibinfo{publisher}{ACM},
  \bibinfo{pages}{1–18}.
\newblock
\href{https://doi.org/10.1145/3613904.3642919}{doi:\nolinkurl{10.1145/3613904.3642919}}


\bibitem[Zhang et~al\mbox{.}(2026)]%
        {zhang2026thinkstrokespixelsprocessdriven}
\bibfield{author}{\bibinfo{person}{Lei Zhang}, \bibinfo{person}{Junjiao Tian},
  \bibinfo{person}{Zhipeng Fan}, \bibinfo{person}{Kunpeng Li},
  \bibinfo{person}{Jialiang Wang}, \bibinfo{person}{Weifeng Chen},
  \bibinfo{person}{Markos Georgopoulos}, \bibinfo{person}{Felix Juefei-Xu},
  \bibinfo{person}{Yuxiang Bao}, \bibinfo{person}{Julian McAuley},
  \bibinfo{person}{Manling Li}, {and} \bibinfo{person}{Zecheng He}.}
  \bibinfo{year}{2026}\natexlab{}.
\newblock \bibinfo{title}{Think in Strokes, Not Pixels: Process-Driven Image
  Generation via Interleaved Reasoning}.
\newblock
\showeprint[arxiv]{2604.04746}~[cs.CV]
\urldef\tempurl%
\url{https://arxiv.org/abs/2604.04746}
\showURL{%
\tempurl}


\bibitem[Zhang et~al\mbox{.}(2023)]%
        {zhang2023planninglargelanguagemodels}
\bibfield{author}{\bibinfo{person}{Shun Zhang}, \bibinfo{person}{Zhenfang
  Chen}, \bibinfo{person}{Yikang Shen}, \bibinfo{person}{Mingyu Ding},
  \bibinfo{person}{Joshua~B. Tenenbaum}, {and} \bibinfo{person}{Chuang Gan}.}
  \bibinfo{year}{2023}\natexlab{}.
\newblock \showarticletitle{Planning with Large Language Models for Code
  Generation}. In \bibinfo{booktitle}{\emph{The Eleventh International
  Conference on Learning Representations}}.
\newblock
\urldef\tempurl%
\url{https://openreview.net/forum?id=Lr8cOOtYbfL}
\showURL{%
\tempurl}


\bibitem[Zhang et~al\mbox{.}(2025)]%
        {zhang2024sketchguidedsceneimagegeneration}
\bibfield{author}{\bibinfo{person}{Tianyu Zhang}, \bibinfo{person}{Xiaoxuan
  Xie}, \bibinfo{person}{Xusheng Du}, {and} \bibinfo{person}{Haoran Xie}.}
  \bibinfo{year}{2025}\natexlab{}.
\newblock \showarticletitle{Sketch-guided scene image generation with diffusion
  model}.
\newblock \bibinfo{journal}{\emph{Comput. Graph.}} \bibinfo{volume}{129},
  \bibinfo{number}{C} (\bibinfo{date}{June} \bibinfo{year}{2025}),
  \bibinfo{numpages}{10}~pages.
\newblock
\showISSN{0097-8493}
\href{https://doi.org/10.1016/j.cag.2025.104226}{doi:\nolinkurl{10.1016/j.cag.2025.104226}}


\end{thebibliography}
